\documentclass[twocolumn]{emulateapj}
\usepackage{subfigure}
\begin{document}

\title{Exploring atmospheres of hot mini-Neptunes and extrasolar giant planets orbiting different stars with application to HD 97658b, WASP-12b, CoRoT-2b, XO-1b and HD 189733b}

\author{Y. Miguel}
\affil{Max Planck Institut f\"ur Astronomie, K\"onigstuhl 17, 69117, Heidelberg, Germany}
\email{miguel@mpia.de}
\author{L. Kaltenegger} 
\affil{Max Planck Institut f\"ur Astronomie, K\"onigstuhl 17, 69117, Heidelberg, Germany\\
Harvard Smithsonian Center for Astrophysics, 60 Garden St., 02138 MA,Cambridge, USA}

\begin{abstract}

We calculated an atmospheric grid for hot mini-Neptune and giant exoplanets, that links astrophysical observable parameters- orbital distance and stellar type- with the chemical atmospheric species expected. The grid can be applied to current and future observations to characterize exoplanet atmospheres and serves as a reference to interpret atmospheric retrieval analysis results. To build the grid, we developed a 1D code for calculating the atmospheric thermal structure and link it to a photochemical model that includes disequilibrium chemistry (molecular diffusion, vertical mixing and photochemistry). We compare thermal profiles and atmospheric composition of planets at different semimajor axis (0.01$\leq$a$\leq$0.1AU) orbiting F, G, K and M stars. Temperature and UV flux affect chemical species in the atmosphere. We explore which effects are due to temperature and which due to stellar characteristics, showing the species most affected in each case. CH$_4$ and H$_2$O are the most sensitive to UV flux, H displaces H$_2$ as the most abundant gas in the upper atmosphere for planets receiving a high UV flux. CH$_4$ is more abundant for cooler planets. We explore vertical mixing, to inform degeneracies on our models and in the resulting spectral observables. For lower pressures observable species like H$_2$O or CO$_2$ can indicate the efficiency of vertical mixing, with larger mixing ratios for a stronger mixing. By establishing the grid, testing the sensitivity of the results and comparing our model to published results, our paper provides a tool to estimate what observations could yield. We apply our model to WASP-12b, CoRoT-2b, XO-1b, HD189733b and HD97658b.

\end{abstract}


\section{Introduction}

Characterization of hot mini-Neptunes (planets with masses larger than 10 M$_{\oplus}$ and a primary atmosphere) and giant planets's atmospheres has been shown for transiting exoplanets, either by secondary eclipse measurements or by transmission spectroscopy (see e.g. \citet{se10}, and references therein). Some chemical species (Na, CO and H$_2$O) have been confirmed to exist in the outer atmosphere of some transiting exoplanets. Sodium lines were detected for HD 209458b \citep{ch02} and HD 189733b \citep{re08}. For HD 209458b carbon monoxide \citep{sn10} and water \citep{de13} were observed. For HD 189733b, CO was also detected \citep{deK13}. Carbon monoxide was also found in the atmosphere of tau Bootis b \citep{br12}, water in XO-1b \citep{de13} and both CO and water vapor was detected in HR 8799c 's atmosphere \citep{ko13}.  These detections along with atmospheric parameter retrieval using broadband photometric data (e.g.,\citet{st10,madu11,li13}) explore the underlying chemistry in hot exoplanet atmospheres, showing a variety of atmospheric composition of hot exoplanets. 

Our paper explores the change in chemical atmospheric species in the observable region of the atmosphere, as a function of astrophysical observable parameters, like orbital distance as well as stellar type. We developed a 1D code for calculating the thermal profile and link it to an atmospheric model (see \citet{ravi}) to explore the atmospheric chemistry of exoplanets. We perform simulations comparing thermochemical equilibrium to disequilibrium chemistry driven by vertical mixing, molecular diffusion and photochemistry, in order to explore the differences in the atmospheres (see, \citet{hu07} for a study on disequilibrium chemistry in brown dwarfs' atmospheres). Most of the photochemical models developed in the last years were applied to specific highly irradiated exoplanets, like the hot giant planet HD 209458b \citep{li03,mo11,ve12} and HD 189733b \citep{li10,mo11,ve12}. Other studies explored CoRoT-2b, XO-1b and WASP-12b \citep{mo12} and WASP-12b \citep{ravi}. Adopting some simplifications, like the imposition of chemical equilibrium at an arbitrary lower boundary and the assumption of constant temperature in the entire atmosphere, \citet{za09a,za09b} used their photochemical kinetic code, to perform the first parameter space explorations in the study of hot giant planet atmospheres. \citet{za09a,za09b} studied photochemistry of sulfur products for planets with different temperatures and showed photochemical models for planets with three different temperatures: 800, 1000 and 1200 K, including variations in the eddy diffusion coefficient and different metallicities (−0.7 $\le$ [M/H] $\le$ 1.7) for a planet located around HD 189733, respectively.  

Photochemical models in the literature, explain the chemistry for specific planets, but do not explore broader trends as a function of stellar flux and semimajor axis of hot planets and their atmospheric chemistry. In this paper we present a parameter space exploration, linking astrophysical observables like orbital distance and stellar spectral type with the change in mixing ratios of chemical species in hot exoplanet atmospheres. Different vertical mixing in the atmospheres is also explored, showing possible degeneracies in the mixing ratios of chemical species. We present a grid of thermal profiles and photochemical mixing ratios of short orbital period H-dominated exoplanet atmospheres, in order to explore what to expect in future observations. Our grid shows planets with semimajor axis of 0.01, 0.015, 0.025, 0.05 and 0.1 AU orbiting different host stars (F, G, K and M). 

Section \ref{model} describes our model, section \ref{results} presents our results - for five known extrasolar planets (\ref{validating}) as well as for the whole grid (\ref{grid}). Section \ref{discussion} discusses the influence of vertical mixing (\ref{Eddy}), clouds and hazes (\ref{clouds}), the profile adopted for the thermal structure (\ref{radiative}) and different elemental abundances (\ref{discussion:clouds}) on our results and section \ref{conclusion} summarizes the paper.

\section{The Model}\label{model}

The study of hot exoplanet atmospheres is a complex problem that requires understanding of the thermal structure, photochemistry and hydrodynamics of the planetary atmosphere. In this paper we developed a 1D model for calculating thermal atmospheric profiles, described in section \ref{T-P} and link it to an atmospheric chemical model developed for studying WASP-12b and differences between the photochemical and thermochemical equilibrium models (see, \citet{ravi}).  The model takes the effect of photochemistry induced by stellar irradiation, considering equilibrium and disequilibrium chemistry as well as vertical diffusion and molecular diffusion into account. In this section we describe the model used for computing exoplanet atmospheres.

\subsection{Thermal profile of the atmosphere}\label{T-P}

For the atmospheric thermal structure, we developed a 1D model for highly irradiated exoplanets, which assumes a gray atmosphere in hydrostatic equilibrium \citep{ha08,gu10,bur10}. We focus on modeling the upper observable region of the atmosphere where the opacities are low and therefore we neglect convection and assume a pure radiative model (see Section \ref{discussion}).  

The upper region of the atmosphere is characterized by temperatures significantly smaller than the effective temperature of the star, therefore we can assume that both radiation fields are decoupled and adopt characteristic mean opacities for each one. Assuming that the thermal contribution from the atmosphere in the visible is negligible, the model employs a two-stream approximation, that treats the absorption of stellar irradiation (shortwave) and the subsequent reradiation (longwave) separately. 

Assuming isotropic radiation, the average temperature profile in the atmosphere as a function of the optical depth ($\tau$) is computed in equation \ref{T} \citep{gu10}:

\begin{displaymath}
T_p^4=\frac{3T_{int}^4}{4}\bigg(\frac{2}{3}+\tau\bigg)+
\end{displaymath}
\begin{equation}\label{T}
\frac{3T_{irr}^4}{4} f \bigg(\frac{2}{3}+\frac{1}{\gamma \sqrt{3}}+\bigg(\frac{\gamma}{\sqrt{3}}-\frac{1}{\gamma \sqrt{3}}\bigg)e^{-\gamma \tau \sqrt{3}}\bigg)
\end{equation}
where $T_{int}$ is the internal temperature of the planet and $T_{irr}$ is the irradiation temperature, that characterizes the flux received from the star and is proportional to the equilibrium temperature ($T_{eq}$). The proportionality constant is $f^{-1/4}$, where $f$ has a value of $\frac{1}{4}$ when assuming global average over the entire planetary surface. $\gamma$ is the greenhouse factor defined as the ratio between mean opacities in the short and long wavelengths. $\gamma<1$ implies that the stellar radiation is more transparent than the emerging one, while a value of $\gamma>1$ implies a more opaque atmosphere. We follow \citet{gu10} and adopt a value of $\gamma=0.6\sqrt{\frac{T_{irr}}{2000}}$ and a thermal mean opacity of $\kappa_{th}=10^{-2}$g/cm$^{2}$ for the hottest planets. For the coolest planets in our grid ($T_{eq}\simeq$1000 K), we use $\gamma$=0.07 and the same value of $\kappa_{th}$ (following \citet{mrf10} for GJ 1214b). 

In our model we divide the atmosphere in 100 layers between $10^{-7}$ and 100 bars, with the high pressure level chosen to show the transition between the region of chemical equilibrium and disequilibrium in the atmosphere. Finally we integrate the equation of hydrostatic equilibrium at each step for the pressure structure calculation using equation \ref{hydro}:

\begin{equation}\label{hydro}
\frac{dP}{dz}=-g\rho
\end{equation}
where the change in pressure ($P$) with heigh ($z$) is equal to the density ($\rho$), multiplied by the gravity of the planet (g, set to be a constant in the thin upper region we are modeling).

\subsection{Atmospheric chemistry calculations}\label{photo}

In our models, we assume a C/O=0.54 and solar elemental abundances, following \citet{as05}. We calculate the equilibrium abundances at each temperature-pressure level, based on the minimization of Gibbs free energy, solving simultaneously the system of chemical equilibrium equations to derive equilibrium mixing ratios as shown in \citet{wh58}. We use those values as initial conditions and recalculate the mixing ratios using photochemistry and disequilibrium chemistry.  

To compute the photochemical mixing ratios, we use a 1D photochemistry code developed by \citet{ravi} for WASP-12b that includes disequilibrium chemistry driven by photochemistry, molecular diffusion and vertical mixing. The photochemistry code employs the reverse Euler method in order to solve the system of differential equations that determines the mixing ratios of all species at different heights in the atmosphere, adopting as lower boundary condition the mixing ratios of the species at thermodynamic equilibrium and zero flux for all the long-lived species at the upper boundary.  Our models include 19 chemical species (O,O(1D), O$_2$, H$_2$O, H, OH, CO$_2$, CO, HCO, H$_2$CO, CH$_4$, CH$_3$, CH$_3$O, CH$_3$OH, CH, CH$_2$, H$_2$COH, C, H$_2$) in 179 reactions. The reaction list adopted in this code allows us to perform simulations in hot regimes, for 700$\lesssim$T$\lesssim$2800 K, note that kinetic rates might not be reliable at temperatures higher than 2800K. Our model grid covers planets with equilibrium temperature from 2800 K to 700 K, cooler planets will be explored in a future paper. The complete list of chemical reactions can be found in \citet{ravi} and for detailed information about the numerical analysis used in the code see \citet{pa01}. 

Vertical mixing time scale is parametrized using the eddy diffusion coefficient ($K_{ZZ}$), as usual in 1D photochemical models \citep{li10,mo11,mo12}.  In our calculations we use a nominal value of  $K_{ZZ}=10^{9}~cm^2/s$, which is a value adopted by \citet{mo12} and explore the dependence of the results on this parameter in section \ref{Eddy}. This coefficient is a big uncertain in 1D models and a different value can lead to significantly different results (see section \ref{Eddy}). 

\section{Results}\label{results}

We first apply our model to five known extrasolar planets in this temperature range and compare our results to the four atmospheric  models published (section \ref{validating}). We show the first model developed specifically for the recently discovered HD 97658b (section \ref{section:HD97658b}). Then, we show a grid of planet models, that can be used for hot planets ( 700$<T_{eq}<$2800 K) with a solar composition H-dominated atmosphere (section \ref{grid}). 

\begin{figure}
\includegraphics[angle=270,scale=.3]{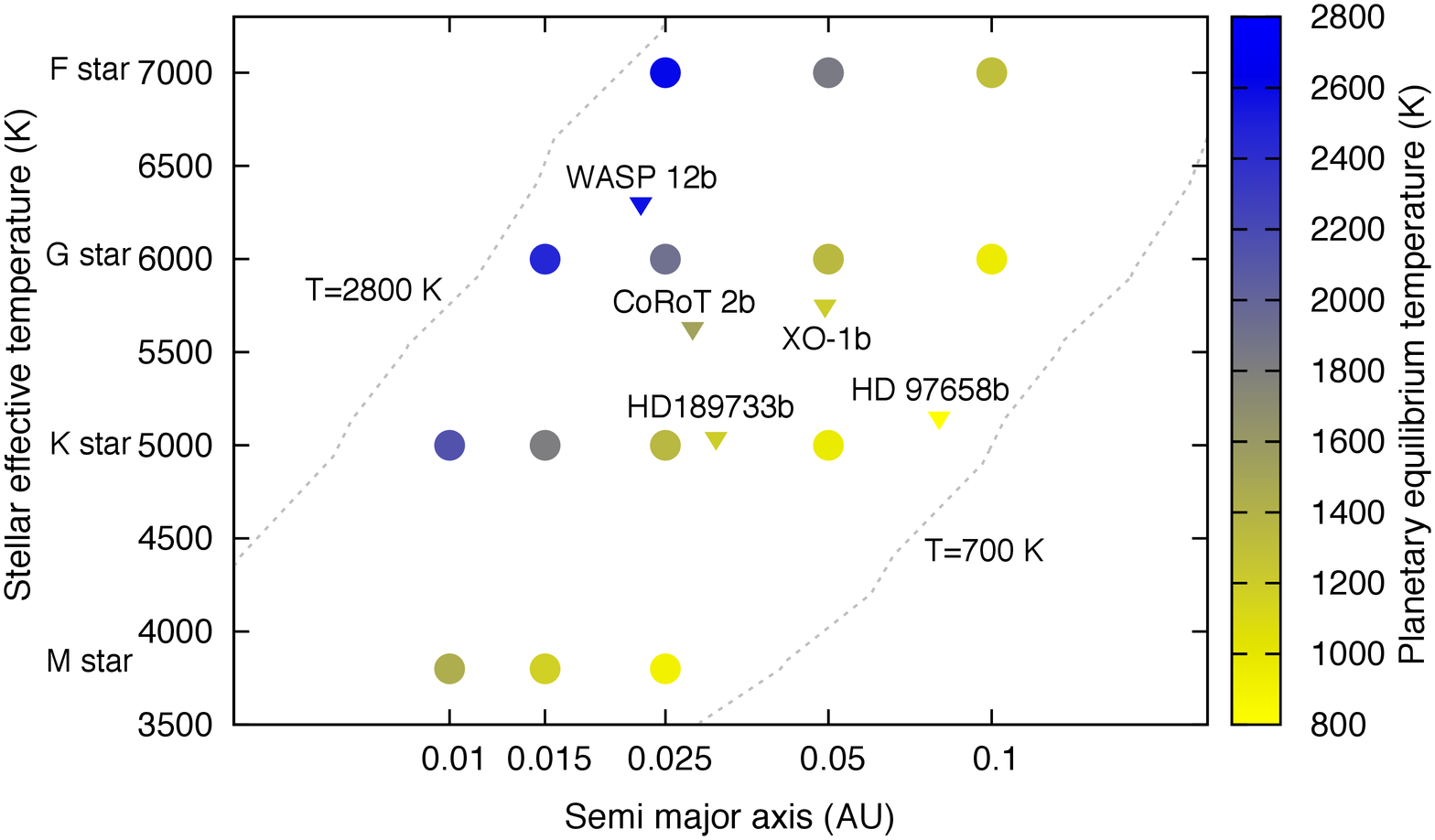}
\caption{Semimajor axis vs. stellar effective temperature for the planets and stars adopted in this paper. The dotted lines show the approximate limits in planetary equilibrium temperature modeled. The dots and triangles represent the planets modeled shown in different color palette according to their equilibrium temperature with hotter planets shown in blue while cooler planets shown in yellow.The 5 known extrasolar planets modeled in this paper are indicated with inverted triangles.}
\label{esquematico}
\end{figure}

The planets modeled in our paper are shown schematically in Figure \ref{esquematico}, which shows the stellar effective temperature vs semimajor axis of the planets assuming a planetary albedo of 0.01 (see e.g., \citet{ro08,bu08,su10}). The inverted triangles indicate the known planets modeled (section \ref{validating}) and the dots the sample planets in the grid (section \ref{grid}), with a color scale according to their equilibrium temperatures ($T_{eq}=T_{eff,\star}\bigg(\frac{(1-A)R_{\star}^2}{4a^2}\bigg)^{\frac{1}{4}}$, where A is the planet albedo, set to 0.01 here). Hotter planets are shown in blue, cooler planets in yellow. Planets with temperatures below 700 K (e.g. a=0.05 and 0.1 around an M star, and a=0.1 AU orbiting a K star) or above ~2800 K (e.g. a=0.01 AU orbiting a G star and a=0.01 and 0.015 around an F star) have equilibrium temperatures which are beyond the temperature range of our model (see section \ref{photo}).  

\subsection{Application of our model to known extrasolar planets}\label{validating}

We calculate thermal and photochemical models for five known exoplanets. We use the four published models to compare our results and validate our model. CoRoT-2b, WASP-12b, XO-1b and HD 189733b fall within the temperature range of our model. Table \ref{tabla-comparacion} shows the main characteristics of these planets and references to recent atmospheric calculations for comparison to our results. Table \ref{tabla-comparacion2} gives the corresponding stellar information.  As seen in the tables, these four planets have semimajor axis between 0.01 and 0.1 AU, orbit stars with effective temperature between 3800 and 7000 K and have a potentially H-dominated atmosphere. Cooler planets like GJ 1214b and GJ 436b were also studied \citep{mrf10,mos13}, but they have cool temperatures that put them outside the range of our study.  We also applied our results to HD 97658b, whose discovery was recently announced \citep{dr13}, where no atmosphere models exist to compare to yet.

\begin{deluxetable*}{lccccc} 
\tabletypesize{\scriptsize}
\tablecolumns{6} 
\tablecaption{Planetary data of the known planets modeled.} 
\tablehead{
\colhead{Planet} &\colhead {a (AU)} & \colhead{M$_p$ (M$_{Jup}$)} &\colhead{R$_p$ (R$_{Jup}$)} &\colhead{ $Log_{10}$(g) } &\colhead{Reference to recent atmospheric calculations}}
\startdata
HD 189733b & 0.031 & 1.14 & 1.138 & 3.341 & \citet{li11a,mo11,ve12}, this paper\\
XO-1b & 0.04928 & 0.918 &  1.206 &  3.211 &  \citet{mo12}, this paper \\
CoRoT-2b & 0.02809 & 3.270 &  1.466 &  3.548 &  \citet{mo12}, this paper \\ 
WASP-12b & 0.02253 & 1.139 &  1.79 &  3.  &  \citet{ravi,mo12}, this paper\\ 
HD 97658b & 0.0796 & 0.0247 & 0.2088 & 3.19 & This paper 
\enddata 
\label{tabla-comparacion}
\end{deluxetable*} 

\begin{deluxetable}{lcr} 
\tabletypesize{\scriptsize}
\tablecolumns{3} 
\tablewidth{0pc} 
\tablecaption{Stellar data of the known planets modeled.} 
\tablehead{
\colhead{Star} &\colhead{R$_{\star}$/R$_{\odot}$} &\colhead{T$_{eff,\star}$ (K)}}
\startdata
HD 189733 & 0.756 & 5040\\
XO-1  &  0.934 &  5750\\
CoRoT-2  &  0.902 &  5630\\
WASP-12  &  1.63 &  6300\\
HD 97658  & 0.703 & 5119 
\enddata 
\label{tabla-comparacion2}
\end{deluxetable}

For the four exoplanets with published photochemical models we use the same stellar flux libraries as in \citet{ravi}, to facilitate the comparison. HD 189733b is orbiting a K2V star, as template we use HD 22049 (following \citet{se03,ravi}). CoRoT-2b, WASP-12b and XO-1b orbit G0 stars, for these planets we use G0 spectra from Pickles’ stellar spectral flux library \citep{pi98}. For HD 97658b, there are no thermal or photochemical models to compare to, therefore we use as input one of the stars used in our grid, with T$_{eff,\star}=$5000 K, R$_{\star}=0.8R_{\odot}$ \citep{ru13}.  

\begin{figure}
\includegraphics[angle=270,scale=.28]{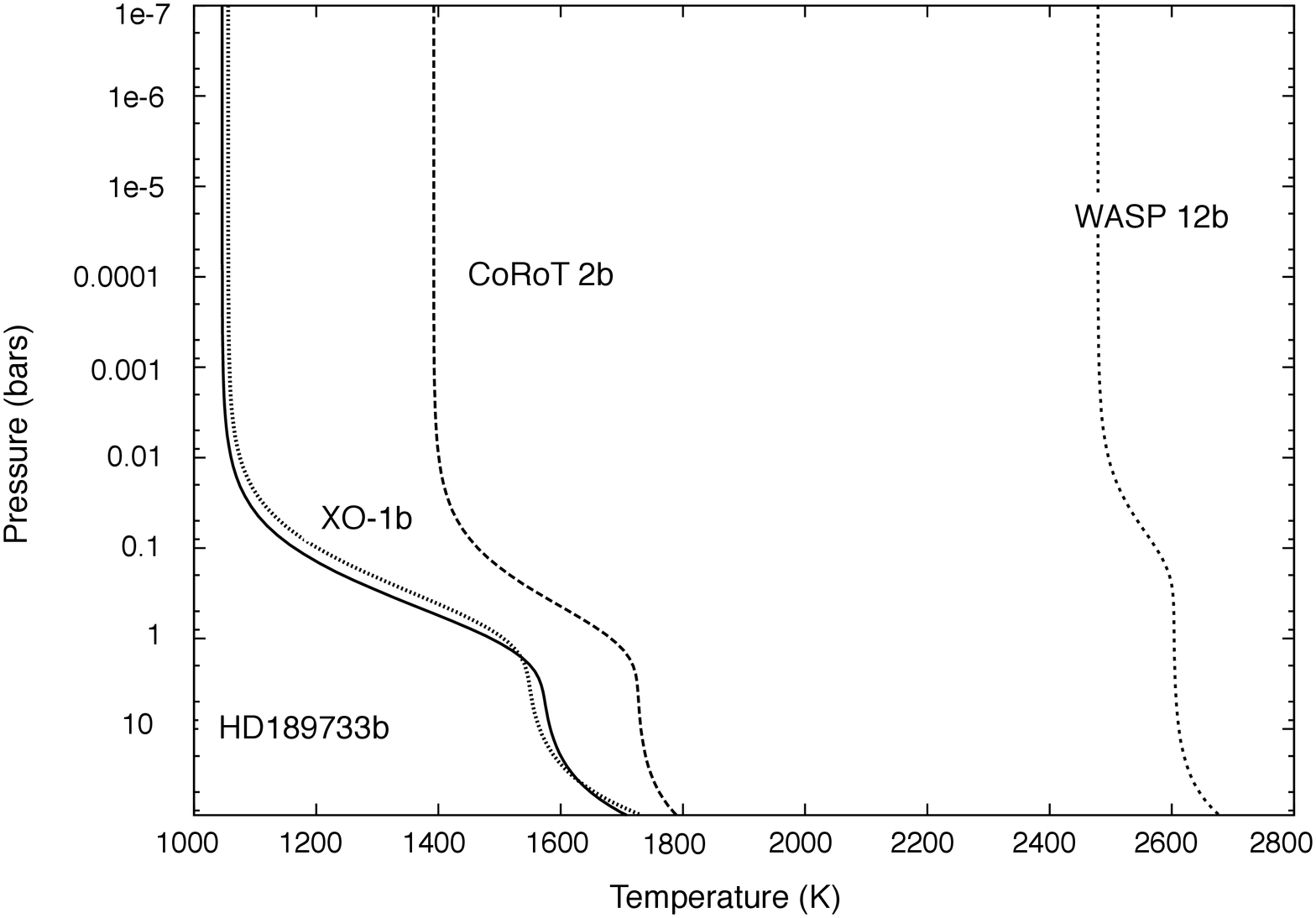}
\caption{Temperature-pressure profile calculated for HD 189733b, XO-1b, CoRoT-2b and WASP-12b.}
\label{comparacion}
\end{figure}

\begin{figure*}
  \begin{center}
\subfigure[]{\label{HD189733b}\includegraphics[angle=270,width=.48\textwidth]{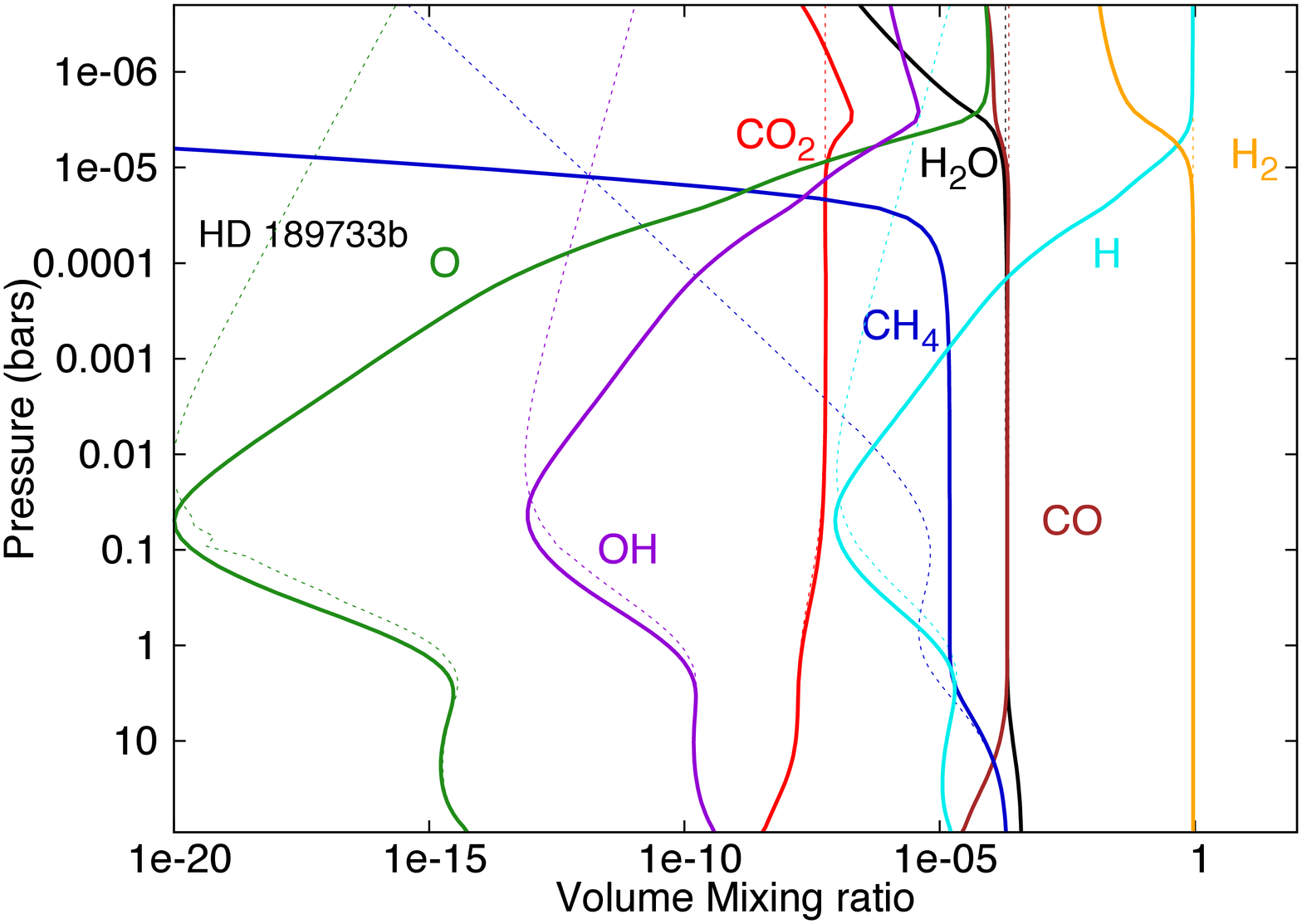}}\subfigure[]{\label{XO-1b}\includegraphics[angle=270,width=.48\textwidth]{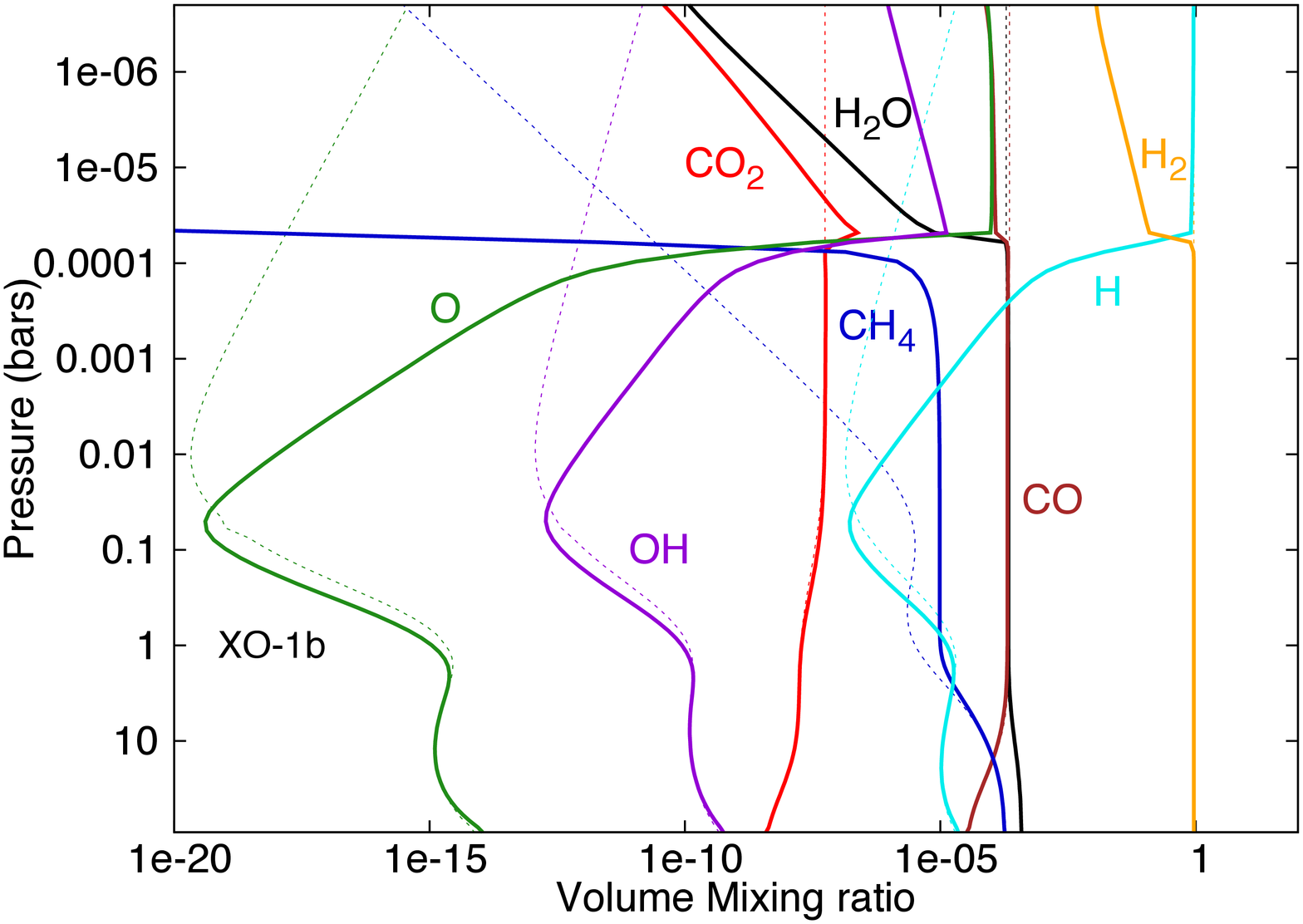}}
\subfigure[]{\label{Corot-2b}\includegraphics[angle=270,width=.48\textwidth]{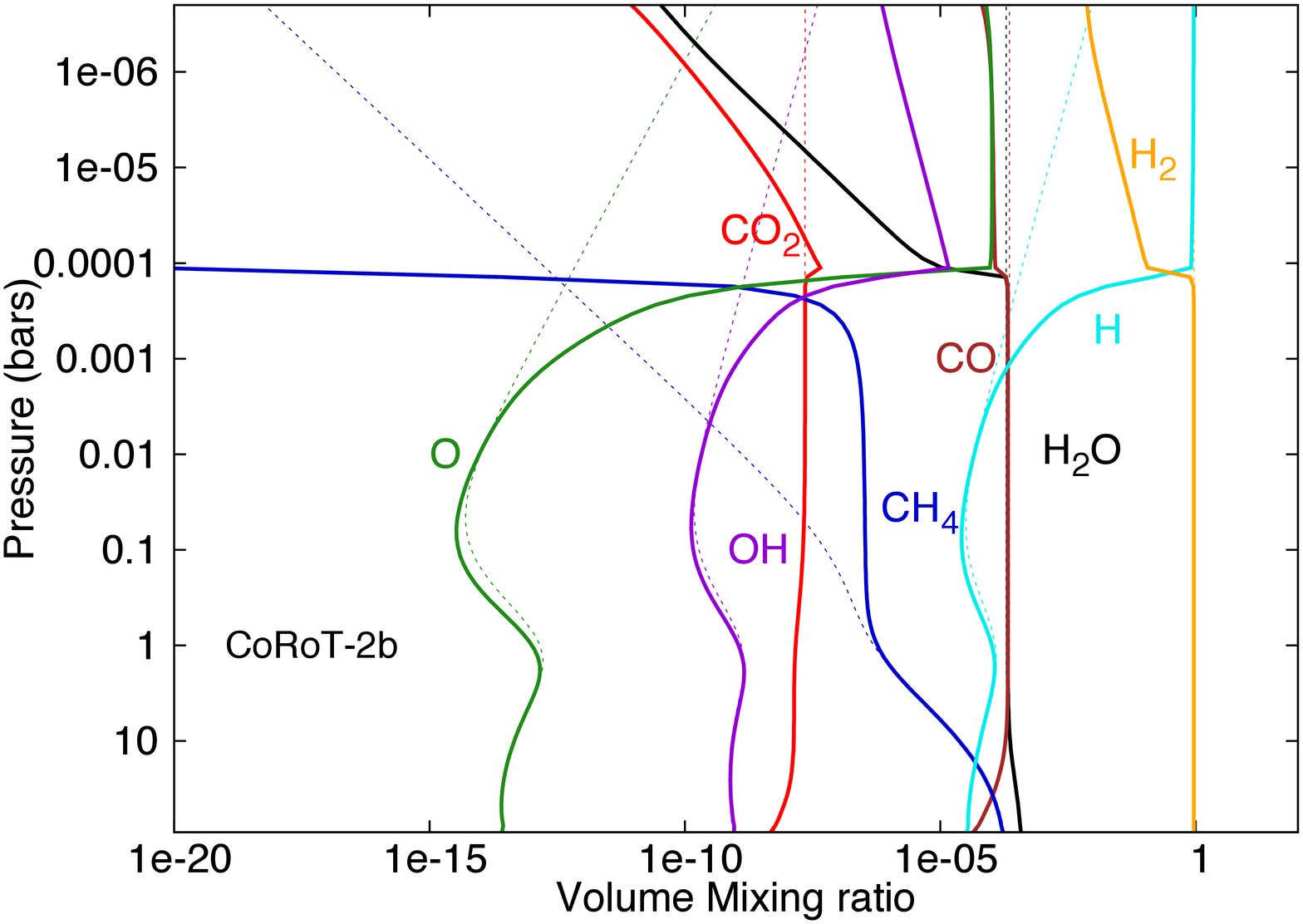}}\subfigure[]{\label{WASP-12b}\includegraphics[angle=270,width=.48\textwidth]{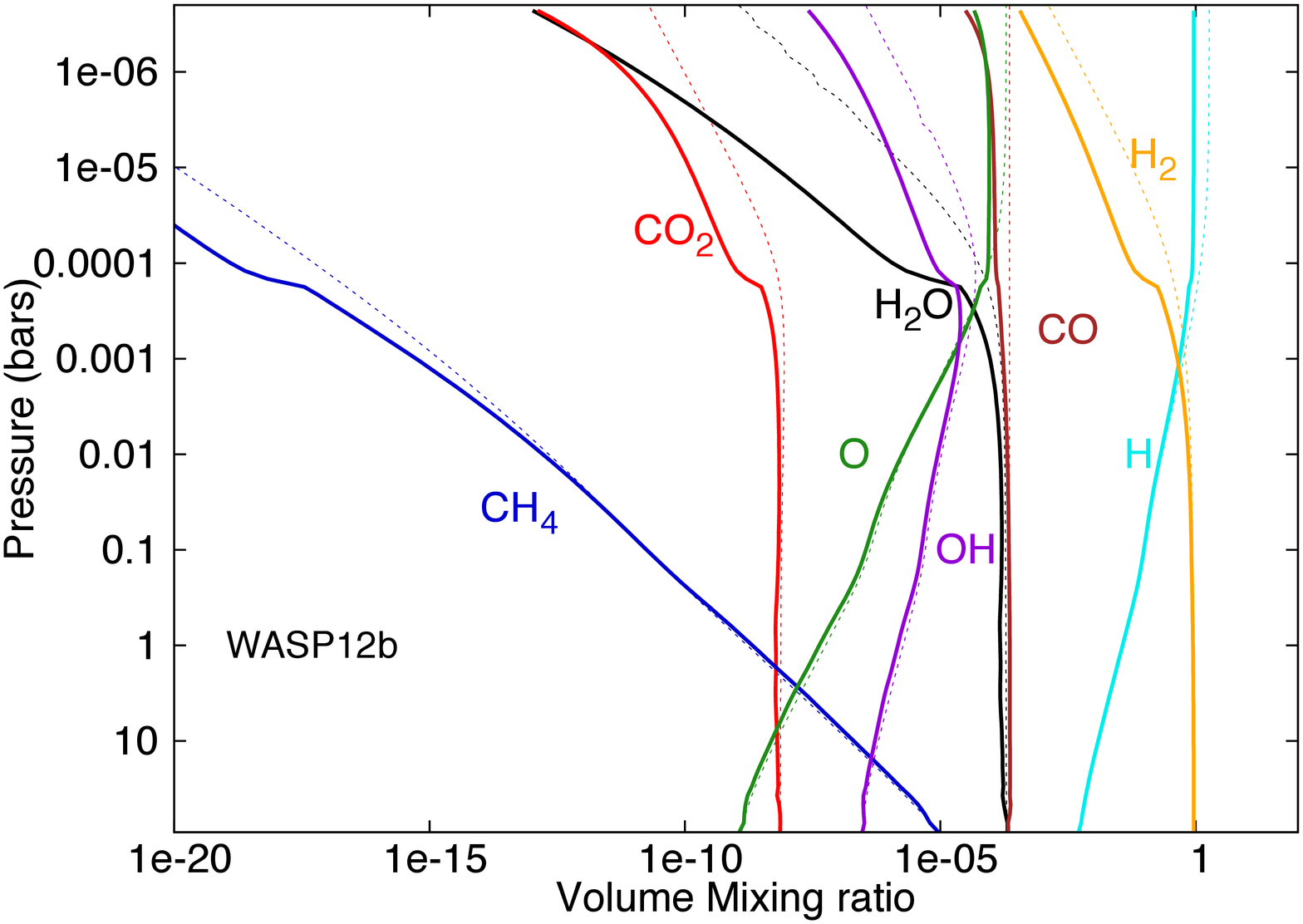}}
 \end{center}
  \caption{Mixing ratios vs. pressures for HD 189733b (Figure \ref{HD189733b}), XO-1b (Figure \ref{XO-1b}), CoRoT-2b (Figure \ref{Corot-2b}) and WASP-12b (Figure \ref{WASP-12b}). The solid lines show the results when disequilibrium chemistry due to vertical mixing and photochemistry is included, while the dotted lines show the mixing ratios in equilibrium.}
  \label{photo-comparacion}
\end{figure*}

Figures \ref{comparacion}  and \ref{photo-comparacion} show the thermal structure and photochemistry profiles, respectively, calculated using the models described in Sections \ref{T-P} and \ref{photo} applied to HD 189733b, XO-1b, CoRoT-2b and WASP-12b. The thermal profile found for HD 97658b is shown in Figure \ref{TP-HD97658b} and its photochemical models are shown in Figure \ref{photo-HD97658b}. In all Figures, the photochemical mixing ratios are shown as solid lines, while the chemical equilibrium values are plotted as dotted lines. Our results agree with published models (see table 1). Small differences in individual values are due to slightly different chemical schemes, thermal profiles, different metallicity, UV fluxes used, and cross sections adopted.

\subsubsection{HD 189733b}

HD 189733b is the most studied planet among the ones analyzed in this paper. As shown in Figure \ref{comparacion}, its thermal model profile is isothermal in the upper atmosphere, with an equilibrium temperature of 1192 K, when adopting a 0.01 Bond albedo, reaching $\sim1500$ K at about 100 bars. Our thermal profile is in good agreement with the profiles calculated by other authors like \citet{bur08} and \citet{sh09} this last authors performed detailed simulations with a global 3D model coupled with a nongray cloud-free radiative transfer code. Other authors \citep{li10,ve12} use a similar T-P profile. 

HD 189733b mixing ratios are shown in Figure \ref{HD189733b}, with CO, H$_2$O and H being the major gases in its atmosphere. The mixing ratios of H$_2$O varies between $10^{-3}$ and $10^{-4}$ until it dissociates at $5\times10^{-6}$ bars. The CH$_4$ mixing ratio varies between $10^{-3}$ and $10^{-8}$ for P$>$ 0.001 bars. OH and O show mixing ratios less than $10^{-9}$ for P$>$0.001 bars. CO mixing ratio vary between $10^{-3}$ and $10^{-4}$. The CO$_2$ mixing ratio is about $\sim10^{-9}$ for values above $5\times10^{-6}$ bars, for lower pressures it increase (due to water photolysis) and then decrease again due to its own dissociation. 

When comparing with other studies we find that CO mixing ratio has similar values in all three comparison studies, but the pressure when the mixing ratio starts to decrease as a consequence of water photolysis depends on the water dissociation level in the atmosphere. For H$_2$O, our model agree with models by \citet{mo11} and \citet{ve12}. Note that \citet{li11a} find the dissociation pressure at a lower pressure of $10^{-8}$. CO$_2$ decreases due to its own photolysis at below $5\times10^{-6}$ bars in our model, in agreement with \citet{mo11} and \citet{ve12}. Note that \citet{li11a} find dissociation at lower pressure of $10^{-8}$ bar for this molecule. We find CH$_4$ levels between $10^{-3}$ and $10^{-5}$ down to pressures of $~10^{-4}$ bars in agreement with \citet{mo11}. Other models find methane levels between $10^{-4}$ to $10^{-5}$ down to pressures of $10^{-8}$ bars \citep{li11a} and $10^{-5}$ and $10^{-6}$ down to pressures of $10^{-4}$ bar \citep{ve12}.

\subsubsection{XO-1b}

The thermal profile for  XO-1b is shown in Figure \ref{comparacion} and presents a isothermal profile with $\sim$1200 K (assuming an albedo of 0.01) in the upper atmosphere and temperatures close to 1800 K in the optically thick region, in agreement with models adopted by \citet{mo12}.

XO-1b photochemical mixing ratios are shown in Figure \ref{XO-1b}, showing H, CO, H$_2$O and CH$_4$ as the major gases in the atmosphere. H$_2$O has a mixing ratio between $10^{-3}$ and $10^{-4}$ until its dissociation, at $\sim5\times10^{-5}$ bars. CO$_2$ shows a mixing ratio between $10^{-8}$ and $10^{-7}$. Water photolysis affects the mixing ratios of other molecules (see section \ref{grid}) and for that reason CO$_2$ has a local maximum at $\sim5\times10^{-5}$ bars, below that pressure it starts to decrease due to its own photolysis. CO, shows a similar behavior and a mixing ratio between $10^{-4}$ and $10^{-3}$ until it starts to decrease at $\sim10^{-7}$ bars. H mixing ratio is also affected by water photolysis, becoming the major gas in the upper atmosphere. CH$_4$  reaches the quench level at a pressure close to 1 bar and starts to decrease at $10^{-4}$ bars. 

Our results agree with \citet{mo12}, with minor differences caused by different metallicities and stellar flux used in both models. While \citet{mo12} adopted a metallicity of Fe/H=$1.5\times$ solar and use the solar spectral flux as an input in their photochemical calculations, we use a solar metallicity and our incoming stellar flux is from a G0 star whose flux in the UV is higher than the Sun (following \citet{ravi}, figure 3).  As a consequence, H$_2$O starts to dissociate at $10^{-6}$ bars for \citet{mo12} and at$\sim5\times10^{-5}$ in our simulations. This change in the H$_2$O dissociation also bring minor changes in the CO$_2$ and CO mixing ratios, but does not influence the overall agreement substantially.

\subsubsection{CoRoT-2b}

CoRoT-2b thermal profile is shown in Figure \ref{comparacion}. This planet has an equilibrium temperature of 1536 K (assuming an albedo of 0.01).  Our thermal profile is similar to retrieval models \citep{madu09,ma11a,madu12}, but differs from the ones adopted by \citep{mo12}. 

Figure \ref{Corot-2b} shows the photochemical and equilibrium mixing ratios of CoRoT-2b. H, CO and H$_2$O are the most abundant species in the model atmosphere. H$_2$O has a mixing ratio between $10^{-3}$ and $10^{-4}$ until it gets photolyzed at $\sim10^{-4}$ bars. Water photolysis affects other molecules like CO$_2$ which has a local maximum due to H$_2$O dissociation at $\sim10^{-4}$ bars. CO mixing ratio is between $10^{-3}$ and $10^{-4}$ in our models and methane quench level is at 1 bar and its abundance is lower than in the previous cases, because this planet is hotter than HD 189733b and XO-1b. 

\citet{mo12} adopted a metallicity of $0.5\times$ solar for the case of solar C/O. They also use as input the solar flux scaled at CoRoT 2b semimajor axis, while we use solar metallicity and the same G0 star as for XO-1b. This difference in the stellar flux explains the difference in the mixing ratios of the major molecules. H$_2$O photolysis occurs at a lower pressure for \citet{mo12}, at $10^{-6}$ bars. This also affects the mixing ratios of other molecules as CO$_2$ whose local maximum consequence of water dissociation occurs at $10^{-7}$ bars in these models. CO and methane  present minor changes. 

\subsubsection{WASP-12b}

WASP-12b is the hottest planet among the known planets modeled in this paper. It has an equilibrium temperature of 2577 K (adopting an albedo of 0.01) and its thermal profile is shown in Figure \ref{comparacion}.  Our thermal profile for WASP-12b agrees with retrieval models \citep{madu09,ma11a,madu12}, but differs from to the ones adopted by other authors \citep{ravi,mo12}. 

Mixing ratios of WASP-12b are shown in Figure \ref{WASP-12b}. H, CO and H$_2$O are the major gases in the atmosphere, while CH$_4$ has a very low mixing ratio as a consequence of the high temperature of the planet and therefore is not expected to have significant abundance at the observable pressure levels.  H$_2$O has a mixing ratio between $10^{-3}$ and $10^{-4}$ for P$>10^{-4}$ bars and start to dissociate at $5\times10^{-4}$ bars. As seen in the previous cases, H$_2$O dissociation also affects the mixing ratios of the rest of the major species in the upper atmosphere. CH$_4$ shows a mixing ratio less than $10^{-5}$,  CO a mixing ratio between $10^{-3}$ and $10^{-4}$ and CO$_2$ a mixing ratio close to $10^{-9}$ until its dissociation at $10^{-4}$ bars.

When comparing with other authors, we notice more differences than the other three cases because the initial assumptions and thermal profiles of \citet{mo12, ravi} are different from ours. To compare between models we keep metallicity and C/O ratio constant in our grid as explained in section \ref{photo}. \citet{mo12} adopted 3 times solar metallicity for this planet which influences in the comparison. Another difference is the eddy diffusion coefficient adopted, which is $K_{ZZ}=10^{10}~cm^2/s$ for \citet{mo12}, and $K_{ZZ}>10^{10}~cm^2/s$ for \citet{ravi} in the upper atmosphere and close to $K_{ZZ}=10^{9}~cm^2/s$ for P$>10^{-2}$ bars. Both values are higher than $K_{ZZ}=10^{9}~cm^2/s$ adopted in our paper. Despite these differences, the CO profile is very similar in the three cases, while there are some differences in the H$_2$O, CH$_4$ and CO$_2$ profiles. H$_2$O has a similar mixing ratio in the three cases, but its dissociation starts at a lower pressure in our model, as a consequence of the hotter thermal planetary profile. The CO$_2$ mixing ratio profile by \citet{ravi} is very similar to the one we find, except in the upper atmosphere, where it is affected by the differences in the H$_2$O photolysis.  \citet{mo12} found a larger mixing ratio of CO$_2$, related to the larger metallicity they adopt in their model. Our methane mixing ratio profile is very similar to the one by \citet{mo12}, being $\sim 10^{-5}$ at 100 bars, and decreasing close to the chemical equilibrium curve until $\sim10^{-4}$ bars. As a consequence of their cooler thermal structure, \citet{ravi} have a larger mixing ratio in the region between $10^{-2}$ to $10^{-4}$ bars, which is close to $10^{-10}$ while \citet{mo12} and our results show mixing ratios of $\sim 10^{-14}$.

\subsubsection{HD 97658b: a planet with an atmosphere in thermochemical disequilibrium}\label{section:HD97658b}

\citet{dr13} announced the discovery of HD 97658b, which is a transiting planet of $7.862~M_{\oplus}$, radius of $2.34~R_{\oplus}$ and whose host star has T=$5119$ K and $R=0.703$ R$_{\odot}$. The mass and radius of HD 97658b implies a low density, which suggests that this planet might have an atmosphere of volatiles. 

\citet{ho12} made some predictions about HD 97658b's atmospheric observables, by assuming similar atmospheric properties as the cool GJ 1214b. They didn't present photochemical models specifically for this planet. Here, we assume that it has an atmosphere with significant H/He (and solar elemental abundances) and model this planet and its potential atmosphere. This is a first model for HD 97658b's atmosphere. Note that other C/O ratios, albedos and metallicities will influence these results. Figure \ref{esquematico} shows the location of this planet in our grid, close to the lower temperature limit. 

\begin{figure}
  \begin{center}
\subfigure[]{\label{TP-HD97658b}\includegraphics[angle=270,width=.45\textwidth]{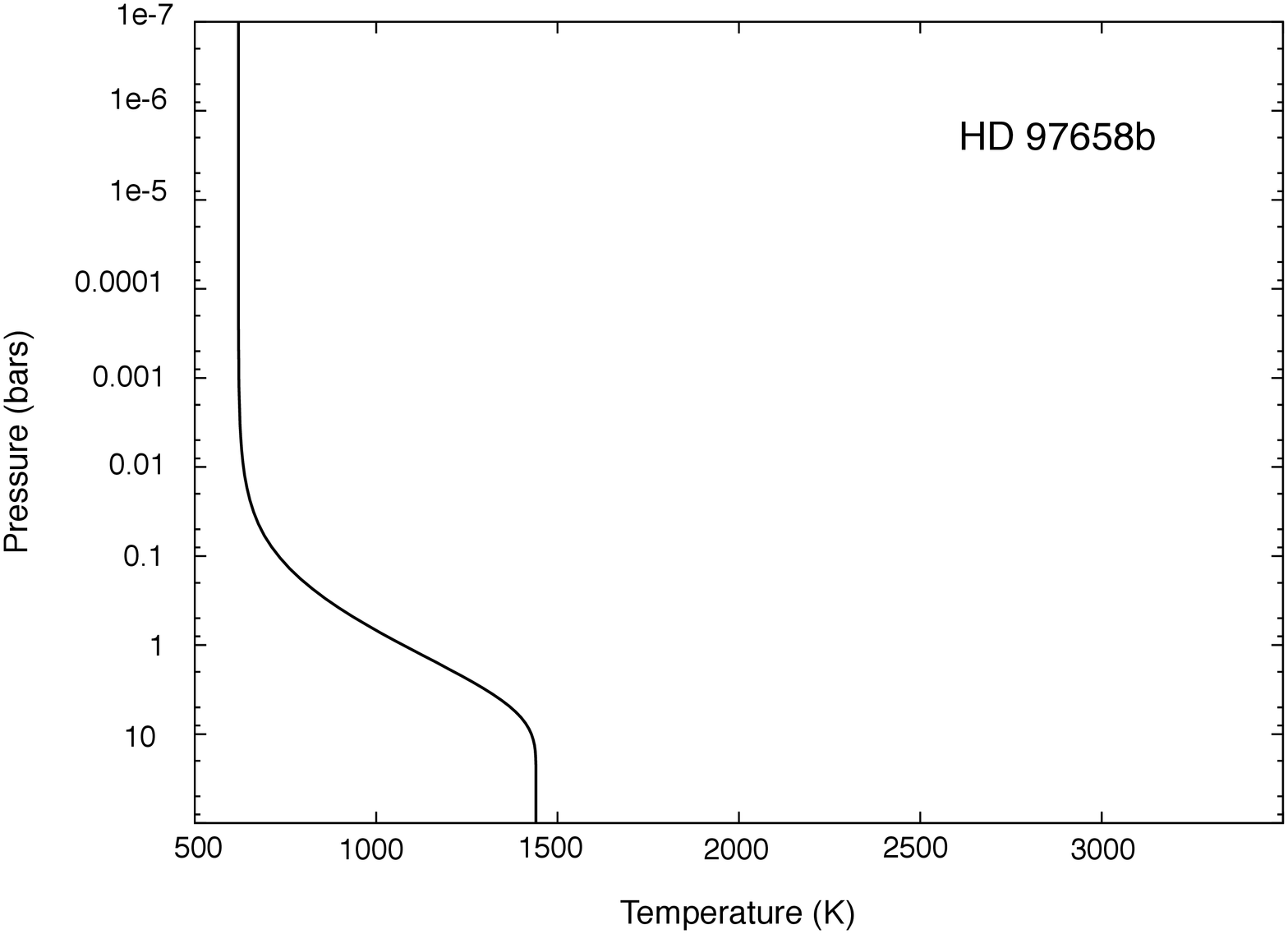}}
\subfigure[]{\label{photo-HD97658b}\includegraphics[angle=270,width=.45\textwidth]{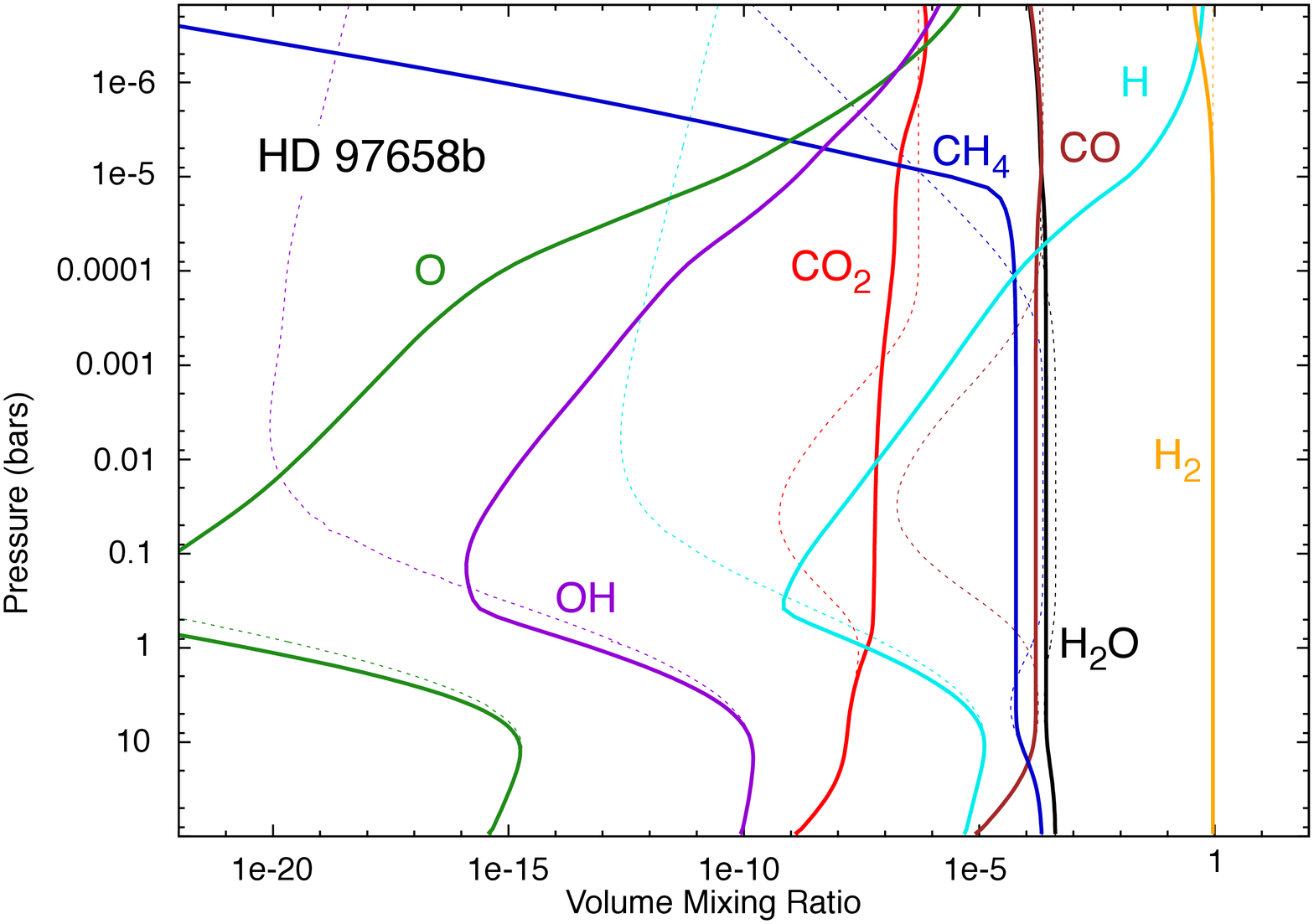}}
 \end{center}
  \caption{Thermal profile (Figure \ref{TP-HD97658b}) and mixing ratios of the major chemical species (Figure \ref{photo-HD97658b}) for HD 97658b . The solid lines show the results when disequilibrium chemistry due to vertical mixing and photochemistry is included, while the dotted lines show the mixing ratios in equilibrium. A value of K$_{ZZ}=10^{9}~cm^2/s$ was adopted.}
\label{HD97658b}
\end{figure}

Figure \ref{TP-HD97658b} shows the thermal profile model for HD 97658b, with an equilibrium temperature of $T_{eq}=731~K$ when adopting a 0.01 Bond albedo. The photochemical mixing ratios for this planet are shown as solid lines in Figure \ref{photo-HD97658b}. As dotted lines we show the equilibrium values, where the model shows a thermochemical disequilibrium for pressures P$<$1 bar. Our model shows that H$_2$, H$_2$O, CO, H and CH$_4$ are the most abundant gases in HD 97658b's atmosphere. 

H$_2$O has a mixing ratio between $10^{-3}$ and $10^{-4}$, until it starts to dissociate. The planet is relatively far from its host star (a=0.0796 AU), when compared with other planets considered in this paper, receiving a lower UV flux from its host star. As a consequence, H$_2$O starts to dissociate relatively high in the atmosphere ($\sim10^{-6}$ bars) and its dissociation is not as efficient as for planets  closer to their host stars. 

Since this is a cool planet, CH$_4$ abundance is relatively high, with a mixing ratio between $10^{-4}$ and $10^{-5}$, until it starts to decrease at $10^{-5}$ bars. CO$_2$ is relatively abundant, with a relatively constant profile close to $10^{-4}$ down to $10^{-5}$ bars, that increases as a side effect of water photolysis (see \ref{grid}). CO has a mixing ratio  between $10^{-3}$ and $10^{-4}$ in all the pressure range explored. H replaces H$_2$ as the major species in its atmosphere at P$<5\times10^{-6}$ bars. 

We adopted an eddy diffusion coefficient of $K_{ZZ}=10^{9}~cm^2/s$ here. Section  \ref{Eddy} discusses how different values influence the chemical abundance of major species in the upper atmosphere.

\subsection{Grid of planetary atmospheres between 0.01 and 0.1 AU around MKGF host stars}\label{grid}

We use our model to build a grid of mini-Neptune and giant planet atmospheres to explore the characteristics of a wide range of exoplanets. This grid links astrophysical observables with exoplanets atmospheric composition and can be used as a reference for current and future observations and retrieval analysis. 

\begin{figure}
\includegraphics[angle=270,scale=.3]{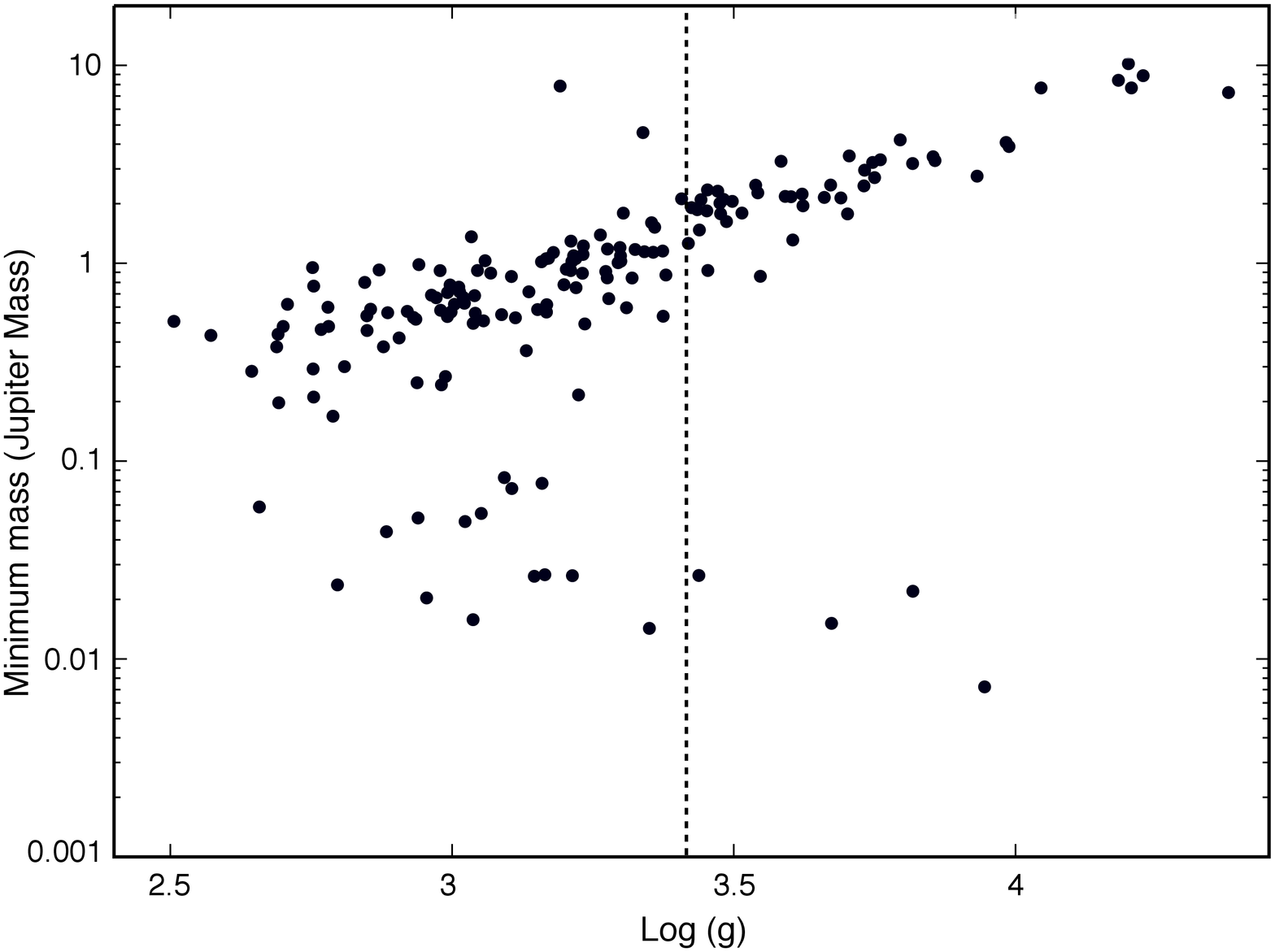}
\caption{Planetary mass vs. surface gravity for all the planets with calculated values in the literature (http://exoplanets.org). The dotted line shows the intermediate value (log(g)=3.4) adopted in our grid.}
\label{gravity}
\end{figure}

This grid is applicable to any hot exoplanet ( 700$<$T$<$2800 K) with H-dominated atmosphere of solar composition and a surface gravity of $log(g)\simeq3.4$. This value was chosen as a mean value that can be applied to model mini-Neptunes or giant planets, as shown in Figure \ref{gravity}, where the surface gravity is shown as a function of the planetary mass\footnote{Figure \ref{gravity} was made using the data published in http://exoplanets.org/ \citep{wr11}}. 

We assume that the primary atmosphere is mainly composed of H and He, with elemental abundances of solar composition. Our grid models the thermal profile and mixing ratios of planets between 700 K and 2800 K, what corresponds to planets located at 0.01, 0.015, 0.025, 0.05 and 0.1 AU of their hosts stars, assuming a planetary albedo of 0.01. 

We use the stellar fluxes for an F (T$_{eff,\star}=$7000 K, R$_{\star}=1.5R_{\odot}$), G (T$_{eff,\star}=$6000 K, R$_{\star}=1.1R_{\odot}$) and K stars (T$_{eff,\star}=$5000 K, R$_{\star}=0.8R_{\odot}$) calculated by \citet{ru13}, based on stellar models taken from the ATLAS synthetic spectra \citep{ku79} combined with UV observations from the International Ultraviolet Explorer (IUE) archive \footnote{http://archive.stsci.edu/iue/}. For the M star (T$_{eff,\star}=$3800 K, R$_{\star}=0.62R_{\odot}$), we use the emission flux of an inactive M star from the dust model developed by \citet{all01}. We use the luminosity of a main sequence star based on its effective temperature and calculated the flux received at the top of the planet atmosphere using the inverse square law of the distance to get the correct flux for each planet's semimajor axis. Note that the inactive Mstar model we use assumes no chromospheric activity \citep{all01}, what might overestimate the differences between active and inactive stars if some chromospheric activity is present in realistic inactive Mstars (see also \citet{se13}). Many main sequence M stars present strong chromospheric activity that produces high-energy radiation. We will address the effect of activity on a planet's spectra in a future paper. 

\subsubsection{Grid of thermal profiles}

\begin{figure*}
\begin{center}
\includegraphics[angle=270,scale=.6]{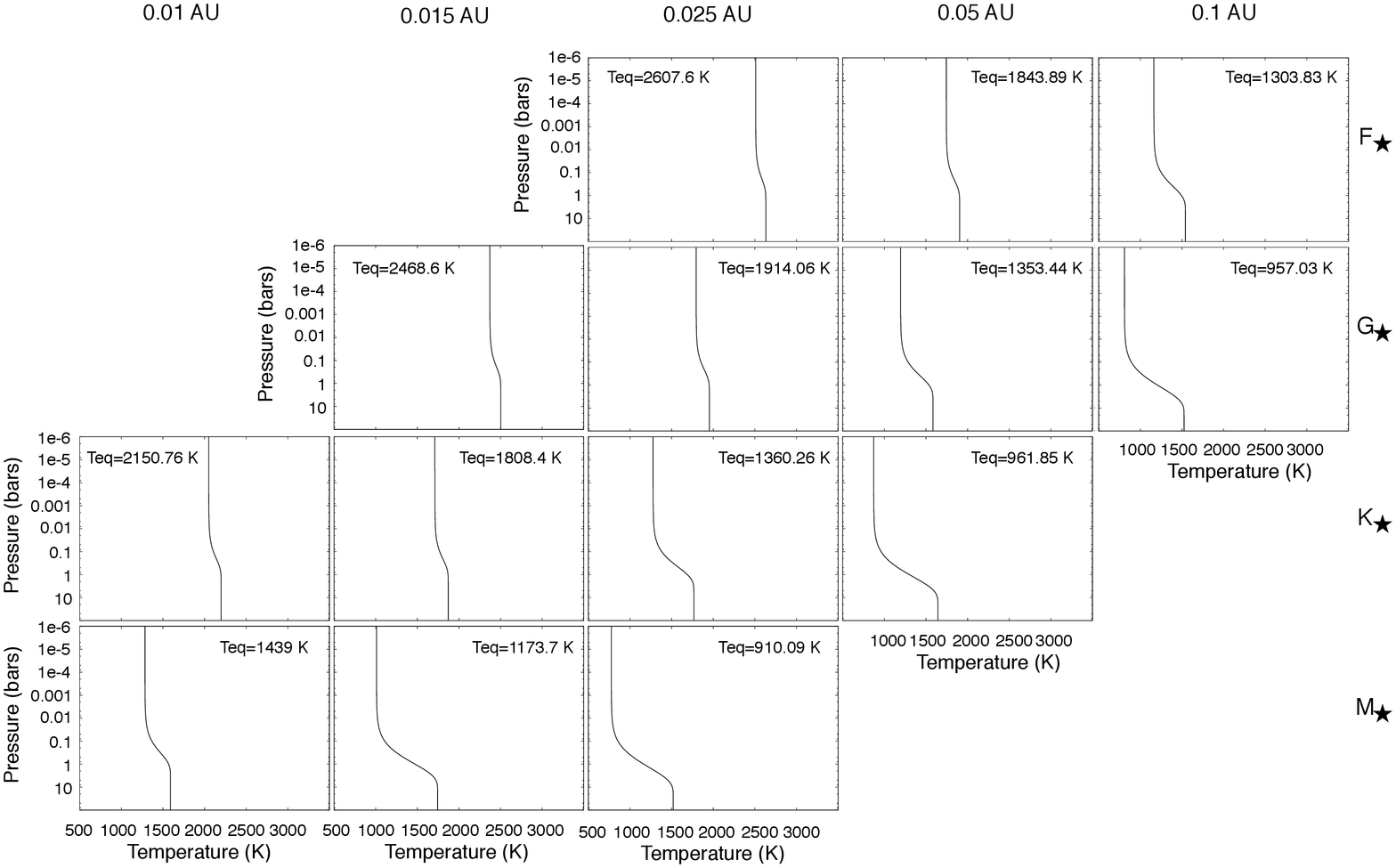}
\caption{Grid of planetary temperature-pressure profiles for H-dominated planets as a function of their semimajor axis. The 5 columns present the distance of the planet from their host star for 0.01, 0.015, 0.05, 0.025, and 0.1AU, from left to right respectively. The different rows show the model results for planets orbiting different host stars. Planets around inactive F (T$_{eff,\star}=$7000 K), G (T$_{eff,\star}=$6000 K), K (T$_{eff,\star}=$5000 K) and M (T$_{eff,\star}=$3800 K) stars are shown in rows 1, 2, 3 and 4, respectively. An albedo of 0.01 was adopted for all planets to calculate the equilibrium temperature with the individual results shown in each panel.}
\label{grid-T}
\end{center}
\end{figure*}

Figure \ref{grid-T} show the grid of atmospheric temperature vs. pressure. Planetary semimajor axis are indicated at the top and the different host star types are shown on the right side of the grid. As expected, temperatures are cooler when the planets orbit further away from the star (see \citet{spi10} for similar results for planets around a solar-type star), with the planet located at 0.025 AU from a M star being the coolest and the one orbiting a F star at 0.025 AU the hottest planet considered in our grid (assuming the same albedo  of 0.01 for all planets).  Figure \ref{grid-T} shows that the planetary thermal profiles typically show an isothermal layer in the upper atmosphere, which is optically thin and the most relevant part when considering the gases that are observable in the spectra. While the incident stellar flux is absorbed, the atmosphere goes from the optically thin to optically thick regime in a transition region, followed by a final, optically thick, nearly isothermal profile below that. Convection is likely to be the most efficient energy transport mechanism at these depths, which is not consider in this paper (see discussion in Section \ref{radiative}), therefore we do not consider our results below $\sim$1 bar reliable. Note that our radiative models give similar results as observations and models by other groups (see section \ref{validating}), even at higher pressures. 

Chemical kinetics timescales required to maintain the species in equilibrium with each other are shorter than vertical mixing for high temperatures and pressures. Therefore, chemical equilibrium is reached faster at high pressures, deep in the atmosphere of these gaseous planets. Chemical disequilibrium dominates the upper atmosphere of highly irradiated planets, where densities are low and disequilibrium processes have shorter timescales. Furthermore, in highly irradiated atmospheres stellar UV radiation photodissociate molecules in the upper atmosphere, causing escape and recombination of atoms. 

\subsubsection{The effect of UV flux only on a exoplanet's atmosphere}

\begin{figure*}
\begin{center}
\includegraphics[scale=.6]{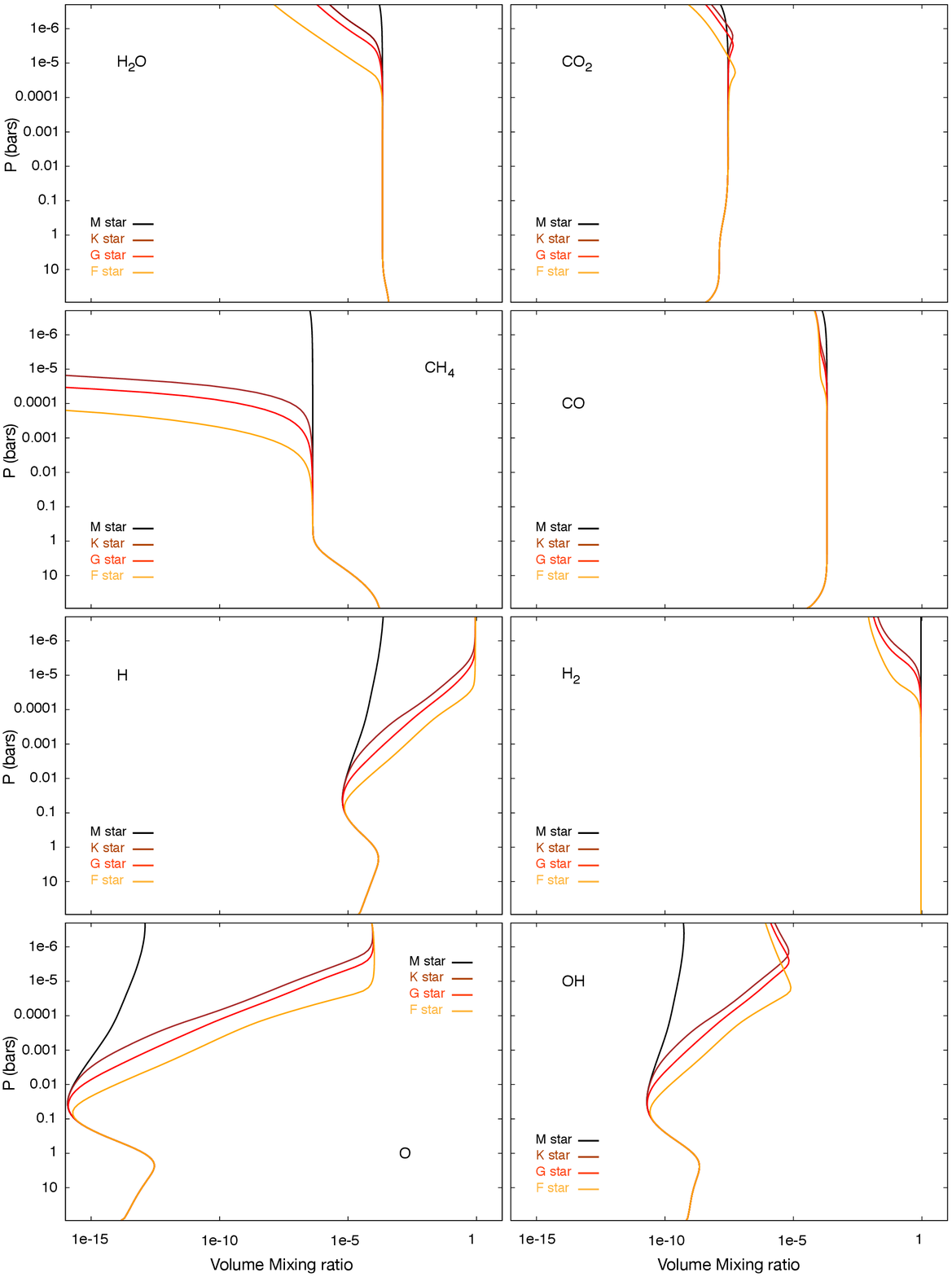}
\caption{Pressure vs. volume mixing ratios for different chemical species. To study the effect of stellar UV flux only, we use the same thermal profile for a planet located at 0.025 AU from a K star (row 3, column 3, in Fig. 6), but change the stellar UV fluxes according to the stellar type: M (black), K (brown), G (red) and F (orange) stars. We adopted  K$_{ZZ}=10^{9}~cm^2/s$.}
\label{SameTP}
\end{center}
\end{figure*}

In order to study the effect of stellar flux alone on photochemical mixing ratios in the atmosphere, we performed simulations using the same thermal profile and vertical mixing, but adopting different stellar fluxes. Figure \ref{SameTP} shows the mixing ratios of different species as a function of pressure for planets with the same thermal structure and eddy coefficient (K$_{ZZ}=10^9cm^2/s$), but different stellar fluxes. For the thermal structure, we use the temperature-pressure profile of a planet located at 0.025 AU of a K star as an example (row 3, column 3 in Figure \ref{grid-T}), the stellar fluxes are the same adopted in the grid calculations. 

The different photochemical mixing ratios found in each case are shown in different colors in Figure \ref{SameTP}, orange (F star), red (G star), brown (K star) and black (M star). H$_2$O and CH$_4$ are the species most affected by the stellar UV flux and the resulting photochemistry. The effect of vertical mixing and photochemistry compete in the atmosphere. Here we investigate the effects of photodissociation, the effect of vertical mixing is explored in section \ref{Eddy}. As expected, at lower pressures in the atmosphere (P$<10^{-4}$ bars) photolysis has shorter timescales than vertical mixing and its effects are even larger when the planet is irradiated by a higher UV flux and therefore mixing ratios change when going from an M to an F star. When larger amount of photons are irradiating the atmosphere, they reach lower altitudes, and therefore dissociation starts at higher pressures. Dissociation of H$_2$O starts at $10^{-4}$ bars when the planet is irradiated by an F star and at $\sim10^{-7}$ bars for an M star case. The differences are even larger for methane, which starts to dissociate at 0.01 bars for the F star case and at $\sim10^{-7}$ bars for the M star. CH$_4$ dissociates creating CH$_3$, H and CH$_2$ and H, therefore methane dissociation creates H.  Large amounts of H are also created due to water photolysis. Dissociation of H$_2$O also creates OH, which reacts with H$_2$,  destroying it. This effect is important in the upper atmosphere, with a H$_2$ mixing ratio of $10^{-2}$ for a planet irradiated by an F star and close to 1 for a planet irradiated by an M star at $\sim 10^{-7}$ bars. Due to this effect, the ratio of H to H$_2$ is higher when going from low to high UV stellar emission (i.e. from M to F stars). CO combines with OH (product of H$_2$O dissociation), forming CO$_2$, that has a local maximum when this happens, at the pressure where water starts to dissociate, then it decreases due to its own dissociation reaching lower mixing ratios for the case of the planet irradiated by the F star. Another effect of water dissociation is the production of O by the reaction between H and OH, which reaches higher mixing ratios in the atmosphere, reaching a mixing ratio close to $10^{-5}$ at different pressures, according to the pressure levels where water starts to dissociate in the atmosphere. Water photolysis dominates the chemistry in the upper atmosphere, being the effects larger when going from the M to the F star. In summary, high stellar UV flux (e.g. F star) leads to increase of mixing ratios of H, OH and O, a small increase in CO$_2$ and the destruction of H$_2$. For planets around stars with low UV fluxes (e.g. inactive M stars) H$_2$, H, H$_2$O, CO and CH$_4$ are the most abundant chemicals in the observable part of the planet's atmosphere.

\subsubsection{Grid of photochemical models}

\begin{figure*}
\begin{center}
\includegraphics[angle=270,scale=.56]{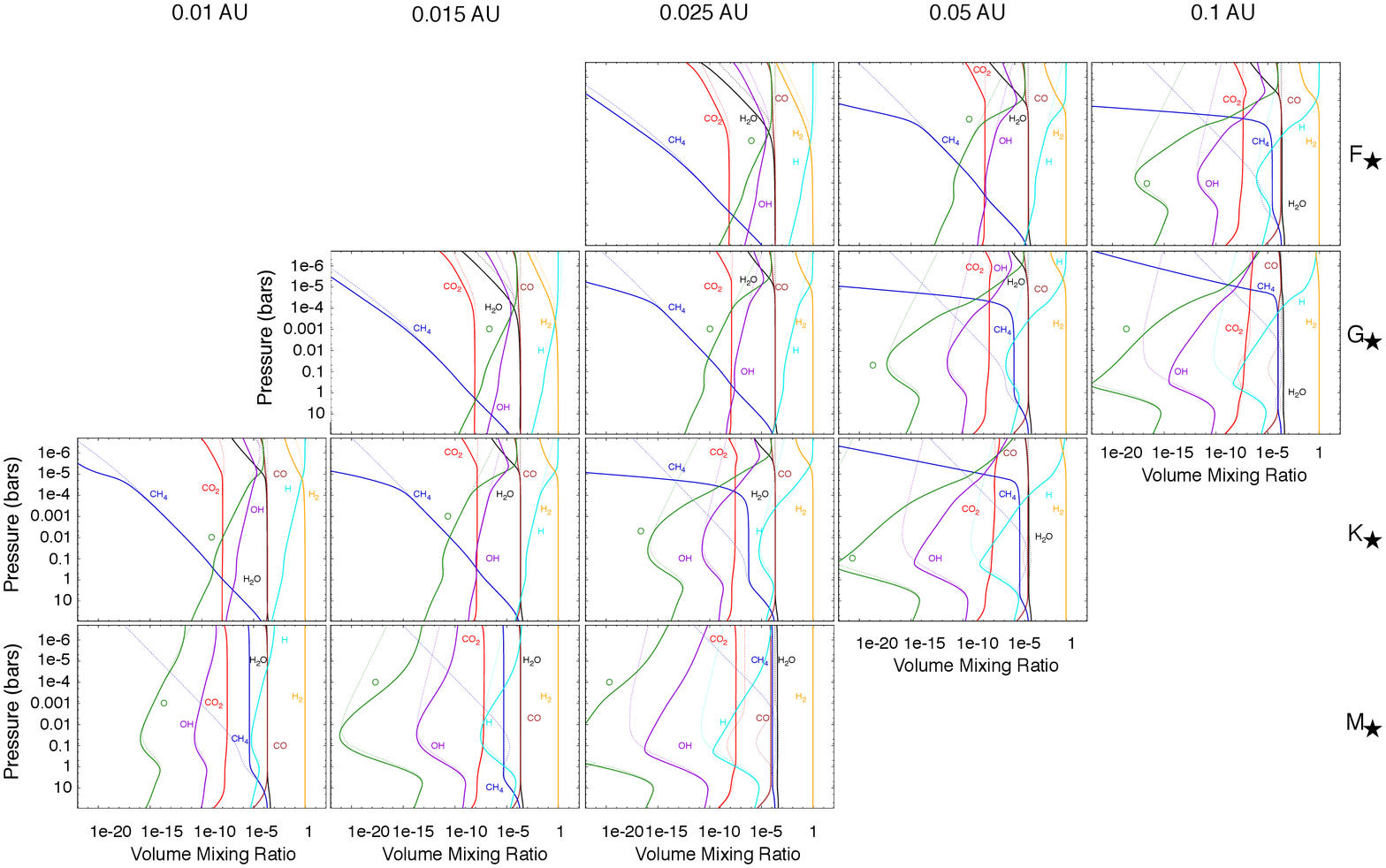}
\caption{Grid of mixing ratios vs pressure for H-dominated planet models orbiting inactive F (T$_{eff,\star}=$7000 K), G (T$_{eff,\star}=$6000 K), K (T$_{eff,\star}=$5000 K) and M (T$_{eff,\star}=$3800 K) stars (1, 2, 3 and 4th row, respectively). The 5 columns present the distance of the planet from their host star for 0.01, 0.015, 0.025, 0.05 and 0.1AU, from left to right respectively. The solid lines show the results when disequilibrium chemistry due to vertical mixing and photochemistry is included, while the dotted lines show the mixing ratios in equilibrium.}
\label{grid-photo}
\end{center}
\end{figure*}

The volume mixing ratios of the major gases in the atmosphere as a function of pressure in the grid are shown in Figure \ref{grid-photo}, where the stellar types of the host star and semimajor axis of the planets are shown on the right and top of the grid, respectively. In this figures, different colors show the major chemical species in dotted lines when they are driven by equilibrium chemistry only and in solid lines when disequilibrium chemistry is also taken into account. As expected at high temperatures and pressures, equilibrium mixing ratios for all species are maintained in the atmosphere (below $\sim$1 bar). Equilibrium levels can be maintained even at lower pressures (P$<10^{-5}$ bars) for the hottest planet around inactive M stars, due to low UV flux emission. It is also shown in the models of the hottest planets in our grid (with equilibrium temperatures close to 2800 K), where the mixing ratios of all chemical species are close to the equilibrium. This is expected because equilibrium is reached faster at high temperatures and pressures, where the chemical equilibrium timescales ($\tau_{eq}$) are shorter than disequilibrium processes ($\tau_{dis,eq}$).  Cooler temperatures and low pressures favor disequilibrium ($\tau_{eq}>\tau_{dis,eq}$), which is relevant for high altitudes in the atmosphere. Therefore, the quench level where $\tau_{eq} \sim \tau_{dis,eq}$, is a function of the temperature, pressure and vertical mixing and will not occur at the same temperature and pressure for all chemical species, because it depends on the time scale for the fastest reactions that produce and destroy the molecule \citep{fe85}. For example, methane departs from equilibrium deeper in the atmosphere than any other chemical species in our models for all the planets of our grid. 

As seen in Figure \ref{SameTP}, high in the atmosphere the UV photons coming from the star are absorbed and the resulting photochemistry dominates the chemical composition of the planet's atmosphere. One of the main effects is the production of large amounts of H and OH due to the photolysis of H$_2$O, followed by a destruction of molecular hydrogen due to its reaction with the produced OH (see e.g. \citet{li03,mo11,ravi}). Therefore, while large amounts of H are created, H$_2$ is destroyed and replaced by H as the main atmospheric species high in the atmosphere. This effect is larger when the photolysis of H$_2$O is larger. Therefore, as seen in Figure \ref{grid-photo}, there is more H and less H$_2$ when the planets are closer to the star, being exposed to more UV radiation.  

CO is the dominant carbon compound in all the planetary atmospheres in the grid. Its abundance is set by thermochemical equilibrium, except high in the atmosphere (P$<10^{-5}$) where it has a local maximum product of water photolysis, except for the hottest planets in our grid, where equilibrium chemistry dominates even at low pressures. 

Methane is more stable at lower temperatures and therefore its abundance increases when going from hot to cooler planets in the grid. In addition, looking at planets with the same semimajor axis but orbiting different stellar types,  CH$_4$ abundance increases also with decreasing stellar temperature, dominated by the effects of decreasing stellar UV. Methane is less efficiently recycled and therefore, even more affected by photodissociation. CH$_4$ is not present in significant amounts high in the atmosphere (above $10^{-5}$ bars) in the planets in our grid, with the exception of the cool planets around the M star, which are exposed to very low UV radiation and are cool enough to maintain a large abundance of methane in the atmosphere. The other exceptions to this are the cases of a higher vertical mixing in the atmosphere (see section \ref{Eddy}). 

As seen in Figure \ref{grid-photo}, the mixing ratio of atomic O increases at high altitudes in all models, due to its production through water photolysis in the atmosphere. Atomic oxygen is also lost due to the backwards reaction, but can be transported up to very low pressures, where it is distributed over large radial distances due to hydrodynamic winds and other processes present in the thermosphere, as explained by several authors (see e.g., \citet{mu09, la09}). Low planetary temperatures (e.g. for large orbital distances) show an increase of CH$_4$. For hot planets  (e.g. with small orbital distances) H, H$_2$ and CO are the most abundant chemicals in the observable part of the planet's atmosphere.

\subsection{Equilibrium vs. Disequilibrium chemistry}

Figures \ref{photo-comparacion}, \ref{photo-HD97658b} and \ref{grid-photo} show the comparison between models that use equilibrium (dotted lines) vs. non equilibrium (solid lines) chemical processes in exoplanet atmospheres. The chemistry in mini-Neptune and giant exoplanet atmospheres is driven by disequilibrium processes (vertical mixing, photochemistry, molecular diffusion) for all planets with temperatures (T $\lessapprox$ 2500K) in our grid, with differences of several orders of magnitude to the corresponding equilibrium chemistry calculations. This occurs at  pressures of P$\lessapprox$ 10 bars, for all the planets in our grid including the known exoplanets HD189733b, XO-1b, CoRoT-2b (Figures \ref{HD189733b}, \ref{XO-1b}, \ref{Corot-2b}) and HD 97658b (Figure \ref{photo-HD97658b}), except for the hottest planets with T $>$ 2500K, where the entire atmosphere is dominated by equilibrium chemistry: the planet at 0.015 AU around an G star, the planet at 0.025AU around an F star. WASP-12b is dominated by equilibrium chemistry for P<0.001 bars (Figure \ref{WASP-12b}). Equilibrium chemistry dominates deeper regions (P$\gtrapprox$ 10 bars) for all planets modeled. Only regions with P$\lessapprox$ 10 bars are observed remotely, therefore, we can not ignore the effects of disequilibrium chemistry when exploring the observable species in these exoplanet atmospheres.

Figure \ref{grid-photo} shows that for all planets in the grid disequilibrium chemistry lowers the abundance of H$_2$, H$_2$O  and CO and increases the abundance of CH$_4$, H, O and OH in the planet's atmosphere. Ignoring disequilibrium chemistry would severely overestimate the abundance of H and H$_2$O and underestimate the abundance of CH$_4$ (for the intermediate atmosphere, it is dissociated in the upper atmosphere), H, O and OH in the observable pressure region for planets with T $\lessapprox$ 2500K.

Figures \ref{photo-comparacion} and \ref{photo-HD97658b} show that using equilibrium chemistry only would overestimate the abundance of H$_2$, H$_2$O and CO for the know exoplanets HD 189733b, XO-1b, CoRoT-2b, WASP-12b and HD 97658b. Note that the effects are smaller for WASP-12b because it is a hotter planet with a T$_{eq}$ of 2577 K (for A=0.01). Especially the abundance of CH$_4$, H$_2$O, H and H$_2$ are critically dependent on disequilibrium chemistry as shown in Figures \ref{photo-comparacion} and \ref{photo-HD97658b}.

\section{Discussion}\label{discussion}

Here we discuss some model parameters that influence our results, vertical mixing in section \ref{Eddy}, clouds and hazes in section \ref{clouds}, our assumption of radiative profile in section \ref{radiative} and elemental abundances in section \ref{discussion:clouds}.

\subsection{Effect of vertical mixing}\label{Eddy}

The eddy diffusion coefficient is included in 1D photochemical models \citep{li10,mo11,mo12}, to represent the vertical mixing processes that occur in planetary atmospheres. This coefficient is hard to determine observationally or through experiments. Some authors use it as a free parameter (see e.g., \citet{mo12}). Others, adopt a coefficient proportional to a globally averaged wind profile from GCM's results, where the proportional factor is the scale height \citep{li10,mo11}. In a recent paper, \citet{pa13} compared 1D and 3D models and derived a $K_{ZZ}$ value that represents the averaged tracer profiles for HD 209458b. These recent results are two orders of magnitude smaller than previous ones obtained using the root mean square of the vertical velocity. These new results suggest that mixing in hot-Jupiter atmospheres is potentially not as strong as previously thought. However, these estimates are an approximation from GCM models calculated for one specific planet and can not be taken as a general result for other planets. 

\begin{figure*}
\begin{center}
\includegraphics[angle=270,scale=.6]{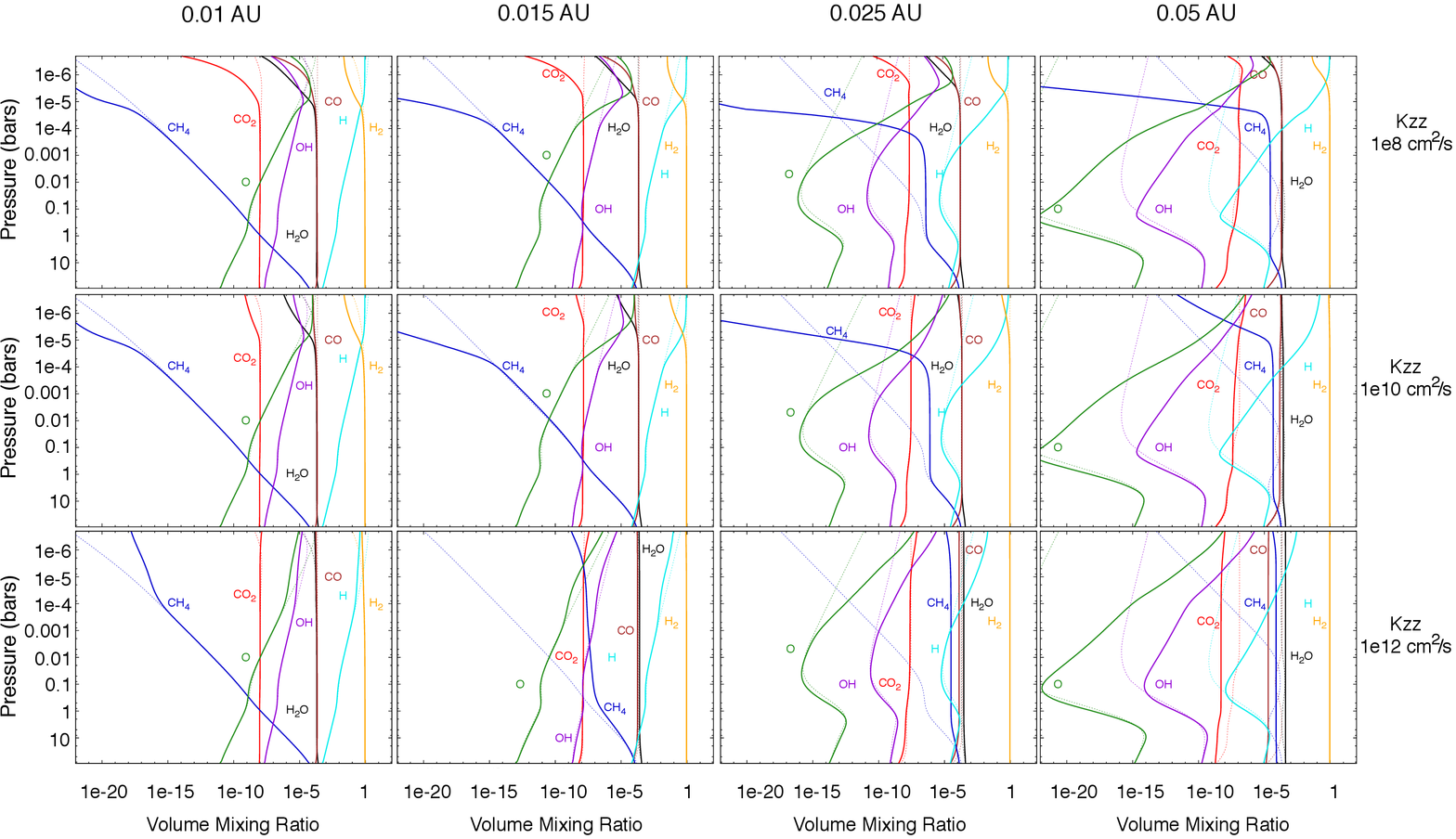}
\caption{Comparison of the mixing ratios for different eddy diffusion coefficients for planets orbiting a K (T$_{eff,\star}=$5000 K) star. The different rows show the mixing ratios as a function of pressure for K$_{ZZ}=10^{8}~cm^2/s$ (top), K$_{ZZ}=10^{10}~cm^2/s$ (middle) and K$_{ZZ}=10^{12}~cm^2/s$ (bottom).  Solid lines show disequilibrium chemistry results, dotted lines show the mixing ratios in equilibrium.}
\label{grid-eddy}
\end{center}
\end{figure*}

Given the uncertainty of $K_{ZZ}$, we model two extreme and one intermediate value here in order to explore the degeneracies in the abundance of observable species on this parameter. Following \citet{pa13} we choose two extreme values that are the minimum and maximum values derived in their parametrization and one intermediate value: $K_{ZZ}=10^8~cm^2/s$, $K_{ZZ}=10^{12}~cm^2/s$ and $K_{ZZ}=10^{10}~cm^2/s$, respectively. Figure \ref{grid-eddy} shows the influence of the eddy diffusion coefficient on the volume mixing ratios as a function of pressure for four planets in our grid located at 0.01, 0.015, 0.025 and 0.05 AU from a K star. Figure \ref{grid-eddy} shows semimajor axis at the top and the $K_{ZZ}$ values on the right side. 

 Figure \ref{grid-eddy} shows that the pressure where the quenched level is reached depends on the vertical mixing. Strong mixing in the atmosphere implies that molecules like CO and CH$_4$ will be quenched at higher pressures, deeper in the atmosphere, than for weaker vertical mixing.  CO$_2$ shows the same behavior for semimajor axis of 0.025 and 0.05 AU.The effects of vertical mixing on the mixing ratios of CO and CH$_4$ was also studied by \citet{vi11}. They analyzed the abundance of CO in the atmosphere of Gliese 229b and CH$_4$ in HD 189733b and found similar results.

In the all cases shown in Figure \ref{grid-eddy}, the mixing is efficient enough to maintain the concentrations of the species well mixed very deep in the atmosphere, where the equilibrium chemistry dominates. Two effects dominate disequilibrium chemistry at high altitudes: vertical mixing and dissociation driven by the UV flux received from the star. Photodissociation of atmospheric molecules due to absorption of energetic photons maintains chemical disequilibrium within the atmosphere, but fast reaction rates in a well mixed atmosphere enable recombination of the molecules split apart by photochemistry. Therefore photochemistry becomes relevant at low pressures in the atmosphere, and it is more efficient when vertical mixing is weak. Figure \ref{grid-eddy} shows that for $K_{ZZ}=10^{12}~cm^2/s$ to $K_{ZZ}=10^{8}~cm^2/s$  (from stronger to weaker vertical mixing) dissociation becomes more efficient. 

Therefore, some molecules that would dissociate at relatively low pressures for a small eddy coefficient, will dissociate at even lower pressures when the mixing is stronger in the atmosphere, becoming one of the major species in the upper atmosphere, like CH$_4$ for planets located at 0.025 and 0.05 AU. Figure \ref{grid-eddy} shows that CH$_4$ is photodissociated at $\sim 10^{-5}$ bars for $K_{ZZ}=10^{8}~cm^2/s$, but has a mixing ratio higher than CO for the same pressure for  $K_{ZZ}=10^{12}~cm^2/s$, showing that the temperature (due to its semimajor axis) is only one of the relevant factors when determining major observable constituents in  an exoplanet's atmosphere. 

CO$_2$ and H$_2$O also show this behavior. A higher dissociation of H$_2$O molecules implies less H$_2$ and more H in the atmosphere, what is shown in Fig. \ref{grid-eddy} when going from stronger to weaker eddy diffusion coefficients. H$_2$ is the major gas in the atmosphere for $K_{ZZ}=10^{12}~cm^2/s$. H becomes the dominant gas for $K_{ZZ}=10^{8}~cm^2/s$, which could be relevant for atmospheric escape studies in hot exoplanets (e.g., \citet{lam13,ku13}). Therefore quantitative measurements from observations of these gases in the atmospheres of hot extrasolar giant planets and mini-Neptunes are needed to constrain the value of the vertical mixing in hot exoplanets atmospheres.   

\subsection{Clouds and hazes} \label{clouds}

We do not consider the existence of atmospheric condensates in our models. The presence of clouds and atmospheric hazes can lead to variations in the opacities, that can alter the thermal profile in the atmosphere. The presence of clouds were inferred for Kepler-7b \citep{dem13} and observations made with the \textit{Spitzer} space telescope of the dayside atmospheres of some hot-Jupiters that presented an excess of emission that was interpretated as thermal inversions in the atmosphere (e.g.\citet{burr07,knu08}). We do not include thermal inversions in the temperature and profiles presented, in order to limit the number of free parameters we explore in this paper. An exploration of possible inversions in these exoplanet atmospheres will be address in a future paper.  

\subsection{Radiative profile} \label{radiative}

Our assumption of a pure radiative model in planets with an extended hydrogen and helium atmosphere is supported by models by several groups \citep{gu96,gs02}, which show that highly irradiated hot-Jupiters present an extended radiative layer. Note that models by \citet{ddl08}, showed that in some cases, the night side of planets might have a convective layer even at low optical depths. Hotter planets may also exhibit weaker global energy redistribution giving them stronger day/night temperature contrasts (e.g.,\citet{cow11}), not considered in our model grid. 

\subsection{Different elemental abundances}\label{discussion:clouds}

The atmospheric elemental abundances of hot exoplanets may differ from that of their host stars. It is an unknown and therefore another parameter to explore using atmospheric models. Recent studies have suggested that lower-mass planets may also have atmospheres highly enriched in heavy elements \citep{for13,mos13}, which is a plausible explanation to the apparent CO-rich, CH$_4$-poor nature of GJ 436b \citep{st10}.  High C/O ratios were suggested in order to explain the abundances detected in WASP-12b \citep{madu11,mo12}, while studies by \citet{cr12,li13}  show that there is no evidence for C/O$\ge1$ in WASP-12b atmosphere.  \citet{mo12}, also explored a range of different C/O ratios ($0.1<$C/O$< 2$) on the atmospheric composition of three exoplanets: CoRoT-2b, XO-1b and HD 189733b, finding the first two consistent with the assumption of high C/O ratios, and HD 189733b consistent with a C/O$\le 1$. These results show that the elemental abundances expected for hot exoplanets is so far not known.

In this paper we focussed on the influence of semimajor axis, stellar flux and vertical mixing on the atmospheric models of gas planets. To limit the parameter space to explore, we adopted solar elemental abundances in the atmospheres. The atmospheric composition and possible deviations of the elemental abundances from solar values on the grid will be explored in a future paper. 

\section{Conclusion}\label{conclusion} 

In this paper we present a grid that links astrophysical observable data (orbital distance and effective stellar temperature) with the atmospheric composition expected in hot-exoplanet atmospheres. The thermal and photochemical atmospheric grid presented in this paper can be applied to current and future planetary observations to inform what observable atmospheric species are expected for H-dominated exoplanet's atmospheres with 700 $<T_{eq}<$2800 K.

We explored the effect of temperature and UV flux linking them to observables like semimajor axis and stellar type in the composition of hot and cool mini-Neptune and giant planet atmospheres models. We used a one dimensional climate and photochemical code to study these effects for five known extrasolar planets (HD97658b, HD 189733b, XO-1b, CoRoT-2b and WASP-12b) as well as provide a grid of planets from 2800 K $<$ $T_{eq} < $700 K. 

We build a grid of planets with different temperatures assuming an albedo of 0.01, calculated based on their semimajor axis (0.01, 0.015, 0.025, 0.05 and 0.1 AU), for a grid of inactive M, K, G and F main sequence stars (T$_{eff,\star}=$3800, 5000, 6000 and 7000 K, respectively). The different semimajor axis dominate thermal profiles of the planets while the stellar spectral energy distribution affects the photochemistry of the atmospheres due to UV radiation received by the planet.

We found that the atmospheres of the hottest planets in the grid are dominated by H, H$_2$, O, H$_2$O and CO, with the last one being the major carbon compound species for all the planets in the grid. For the cooler planets other gases like CH$_4$ become dominant, due to the fact that methane is more stable at lower temperatures. 

Different stellar spectral types supply different levels of UV flux, going from low flux for the inactive M star ($\sim10^3$ photons/s/cm$^2$/\AA$~$ for 1500\AA) to high UV flux for the F star ($\sim10^{10}$ photons/s/cm$^2$/\AA$~$for 1500\AA). This in turn affects the photochemistry in a planet's atmosphere due to the dissociation of molecules occurring down to higher pressure when going from M to F stars. 

Our results show that CH$_4$ and H$_2$O are the most sensitive species to UV flux, and this in turn affects the mixing ratios of other major observable species like CO$_2$, H, H$_2$, OH, having a higher photolysis rate, and therefore a lower mixing ratio for pressures P$\lessapprox10^{-5}$ bars for planets orbiting an F star than an M star. As a result of the photolysis of H$_2$O, and the subsequent destruction of H$_2$ in the reaction with OH, large amounts of H are produced, which replaces H$_2$ as the major gas in the atmosphere, for planets receiving high UV radiation. This effect is important when the photolysis of H$_2$O is large and therefore is higher when going from M to F host stars. 

Vertical mixing is important in exoplanet atmospheres, since different possible values change the mixing ratios of observable gases in the atmospheres. Mixing ratio of CH$_4$ is indicative of the effect of vertical mixing and therefore departure of equilibrium in the atmosphere, specially for pressures close to 1 bar. When looking at lower pressures ($10^{-4}>$P$>10^{-7}$bars), Other species, like H$_2$O, H, H$_2$ or CO$_2$ also indicate the efficiency of vertical mixing in the atmosphere, with larger mixing ratios for stronger mixing in the planet's atmosphere. Dissociation becomes less efficient when going from weak to strong vertical mixing in the atmosphere, which affects the observable atmospheric abundances of most of the species. Measurements of these gases' abundance for short orbital period exoplanets can be used to explore and constrain vertical mixing in these atmospheres. 

\acknowledgments{
We would like to thank Ravi Kumar Kopparapu for in depth discussions. We are grateful to Jonathan Fortney, Ian Crossfield, Andras Zsom, Brad Hansen and Jim Kasting for stimulating discussions and comments that strengthened our paper, as well as Sarah Rugheimer for providing the stellar fluxes. We thank the referee for fruitful suggestions.  LK and YM acknowledge DFG funding ENP Ka 3142/1-1 and the International Space Science Institute ISSI. This research has made use of the Exoplanet Orbit Database and the Exoplanet Data Explorer at exoplanets.org.}

\end{document}